# Dynamics of Nanoscale Phase Decomposition in Laser Ablation

**Authors**

Yanwen Sun[1*‡], Chaobo Chen[2‡], Thies J. Albert[3], Haoyuan Li[1], Mikhail I. Arefev[2], Ying Chen[1], Mike Dunne[1], James M. Glownia[1], Matthias Hoffmann[1], Matthew J. Hurley[4,7], Mianzhen Mo[1], Quynh L. Nguyen[1], Takahiro Sato[1], Sanghoon Song[1], Peihao Sun[5], Mark Sutton[6], Samuel Teitelbaum[7], Antonios S. Valavanis[2], Nan Wang[1], Diling Zhu[1], Leonid V. Zhigilei[2*], Klaus Sokolowski-Tinten[3*]

**Affiliations**

[1]Linac Coherent Light Source, SLAC National Accelerator Laboratory, USA.
[2]Department of Materials Science and Engineering, University of Virginia
[3]Department of Physics, Universität Duisburg-Essen, Germany.
[4]Department of Physics, Stanford University, USA.
[5]Department of Physics, Università degli Studi di Padova, Italy.
[6]Department of Physics, McGill University, Canada.
[7]Department of Physics, Arizona State University, USA.

[‡] These authors contributed equally to this work

[*]To whom correspondence should be addressed; E-mail: yanwen@slac.stanford.edu, lz2n@virginia.edu, klaus.sokolowski-tinten@uni-due.de.

## Abstract

**Femtosecond laser ablation is a process that bears both fundamental physics interest and has wide industrial applications. For decades, the lack of probes on the relevant time and length scales has prevented access to the highly nonequilibrium phase decomposition processes triggered by laser excitation. Enabled by the unprecedented intense femtosecond X-ray pulses delivered by an X-ray free electron laser, we report here results of time-resolved small angle scattering measurements on the dynamics of nanoscale phase decomposition in thin gold films upon femtosecond laser-induced ablation. By analyzing the features imprinted onto the small angle diffraction patterns, the transient heterogeneous density distributions within the ablation plume as obtained from molecular dynamics simulations get direct experimental confirmation.**

## Teaser

The emergence of density heterogeneities in laser ablation is revealed by X-ray probing and explained through atomistic modeling.

# Introduction

Pulsed lasers have become a routine tool for materials processing, synthesis, and analytics. In particular, precise surface texturing and microstructure modification can be achieved with lasers, as well as the synthesis of advanced materials that cannot be obtained by conventional techniques (*1*). Most of these applications involve laser ablation, the removal of macroscopic amounts of material from a laser-irradiated surface.

In this context, ultrashort laser pulses of femtosecond to picosecond duration have demonstrated some remarkable advantages. Surface- and three-dimensional modification/functionalization of materials down to the nanometer scale, pulsed-laser-deposition of extremely smooth, particulate free films and tailored production of nanoparticles represent just a few examples (see (*2*), which provides a comprehensive overview, and references therein).

Besides its importance for applications, the phenomenon of laser ablation provides a fertile playground for the exploration of fundamental questions on the material behavior under conditions of extreme levels of electronic, mechanical, and thermodynamic nonequilibrium. Again, ultrashort laser pulses provide a unique way to induce ablation, as the energy deposition – initially only to the electronic degrees of freedom of the system – is essentially decoupled in time from the subsequent thermal processes. The non-thermal processes caused by strong electronic excitation (*3*), the kinetics of energy transfer from the excited electrons to atomic vibrations, and the subsequent melting and explosive phase decomposition driven by an interplay of extreme thermodynamic (strong superheating) and mechanical (pressure gradients) driving forces (*4*) are the aspects of laser ablation that are actively debated in the research community.

The relevant processes, from the initial energy deposition up to the formation of the final microstructure and/or deposition of ablation products, occur over an extremely large time range from fs to ms (12 orders of magnitude!). Similarly, the spatial scales of the associated structural changes also cover many orders of magnitude – from Å (atomic scale) to hundreds of µm (*e.g.*, diameter of the ablation crater). This massively multi-scale nature of laser ablation poses a major challenge for its experimental investigation, theoretical description, and modeling.

On the experimental side, the discussion of the ablation dynamics is often based on indirect evidence drawn from *ex situ* observations, *e.g.,* analysis of crater shapes, ablation rates as a function of laser parameters such as fluence, pulse duration, and number of applied pulses, as well as size distributions of emitted nanoparticles. Time-resolved studies are scarcer, but provide more direct insights into the ablation dynamics. Of particular relevance to this study are experiments employing femtosecond time-resolved optical microscopy (*5-8*), which have revealed a *universal* ablation behavior at fluences not too far above the ablation threshold, leading to the formation of a characteristic optical interference pattern (Newton rings). These observations provide clear evidence that the ablation plume (i) is largely confined to a region with optically steep (scale length $< \lambda/10$) boundaries and (ii) exhibits low effective absorption (*5).* The extreme steepness of the ablation front towards vacuum has been confirmed by time-resolved interferometric studies (*9*).

From a theoretical/modeling perspective, molecular dynamics (MD) simulations have provided so far the most detailed insight into the microscopic processes underlying laser ablation (*10-14*). In particular, these investigations have revealed that close to the ablation threshold, the material ejection is largely driven by photomechanical effects, where the relaxation of laser-induced stresses leads to the separation (or spallation) of a thin liquid layer from the bulk of the target. This mechanism is not only *universal* (*i.e.*, can occur in almost any material), but could also readily explain the experimental observation of transient Newton rings as the result of interference of light reflected at the spalled layer and the boundary of the non-ablating material (*5, 11, 12, 14*). At higher laser fluences, significantly above the spallation threshold, the simulations reveal the transition to the regime of "phase explosion" (*4,10,12,15*), when the top part of the target is superheated close to the critical temperature and undergoes a rapid (*i.e.* explosive) decomposition into vapor, atomic clusters, and small droplets.

Each regime is characterized by a specific evolution of nanoscale density heterogeneities leading to the material disintegration and ejection, from the nucleation, growth, and coalescence of subsurface nanovoids in the spallation regime (*11-14*) to the rapid volumetric release of vapor in the phase explosion regime (*12, 14*). These initial dynamic processes occurring on the timescale of hundreds of picoseconds set the stage for the long-term evolution of the ablation plume and have direct implications for practically relevant aspects of the ablation process, such as the size distribution of the generated nanoparticles or the structure/morphology of new surfaces produced by ablative micro/nanomachining. Despite the fundamental interest and practical relevance, the direct experimental verification of the computational predictions on the initial stage of laser ablation is still lacking due to the absence of suitable characterization techniques.

The ultrafast optical imaging discussed above provides sufficient temporal resolution for tracking both the initial ablation dynamics and slower processes of ablation plume expansion and target resolidification, but lacks the spatial resolution in the deep sub-µm to nm range required to probe the onset of phase decomposition. X-ray diffraction and scattering techniques at synchrotrons, on the other hand, have been widely used to determine the structure of materials down to atomic resolution. Regarding the application of these techniques to laser ablation, the relatively low X-ray flux per pulse and the long X-ray pulse duration only allowed investigation of slow dynamics, *e.g.*, nanoparticle agglomeration in laser-induced cavitation bubbles in liquids (*16*), or the final ablation product (*17*).

During the past decade, the advent of ultrashort pulsed, short wavelength free electron lasers (XFELs) revolutionized the X-ray probing of structural dynamics by adopting the pump-probe scheme, which is well established in the optical domain: a femtosecond optical laser pumps a sample while a synchronized short-pulse X-ray probe measures the resulting structural changes after a given delay between pump and probe with sub-ps or better resolution (*18*). This scheme has been used to study processes that are also relevant to short pulse laser ablation, namely solid-to-liquid phase transitions following ultrafast laser excitation (*19,20*). However, these studies focused on the structural response on atomic length scales, *i.e.*, the transition from the ordered solid to the disordered liquid, by measuring the transient changes of Bragg diffraction peaks, of the diffuse scattering, and the formation of liquid phase scattering rings. The power of combining the wide- and small-angle X-ray scattering (WAXS and SAXS) probing with atomistic modeling was recently demonstrated in a study of laser fragmentation of colloidal nanoparticles (*21*), where the WAXS signal was used to determine the nanoparticle melting threshold and the SAXS signal provided information on the fragmentation products.

In our work, by utilizing ultrashort, hard X-ray pulses from an XFEL, we apply a combination of time-resolved WAXS and SAXS probing to follow the structural evolution of the material from the initial solid phase into the volatile, ablating state. Our measurements reveal that the material first undergoes a transition to the liquid state, where the loss of long-range order is evidenced by the decay of the wide-angle Bragg peaks, before a dramatic increase of the SAXS signal signifies nanoscale phase decomposition of the superheated and/or overstretched molten phase in the course of ablation. The observed SAXS patterns exhibit a rich, fluence-dependent, and rapidly evolving structure.

Comparing the experimental data to the results of large-scale atomistic simulations, we are able to analyze the complex SAXS signals in terms of the different ablation regimes and the corresponding characteristic pathways of phase decomposition. As such, we arrive at a comprehensive view of the energy deposition and redistribution, the kinetics of the phase transformations, as well as the thermodynamic driving forces behind the spatially heterogeneous and highly nonequilibrium ablation process, which for the first time, is supported by both, direct experimental and modeling data.

## Results

### Experimental results

To trigger the ablation process, 100 nm and 250 nm thick gold (Au) films deposited on arrays of free-standing silicon nitride ($Si_3N_4$) membranes (100 nm thickness) in a silicon wafer frame were irradiated with 50 fs, 800 nm laser pulses and subsequently probed with ultrashort X-ray pulses at 9.5 keV in a normal-incidence transmission geometry (as sketched in Fig. 1 (**A**, **B**)).

The small-angle scattering from the sample was captured by four X-ray area detectors (ePix100 detectors (*22)*; see an example displayed in Fig. 1 (**C**)). The finite size of the detector housing resulted in gaps between the detector sensors and gave rise to "blind zones" shown as dark blue regions in the SAXS pattern. The direct X-ray beam was incident on the lower right detector and blocked by a beam stop right before the detector. The detector pixels in the shadow of the beam stop were also removed from the analysis and colored dark blue in the displayed SAXS patterns. The SAXS image shown in Fig. 1 (**C**) was obtained by averaging over 20 independent measurements performed at a fixed laser fluence of 6.3 J/cm² and an X-ray laser delay of $\delta t$ = 250 ps with respect to the optical laser pulse. Each measurement was performed on a new sample, as detailed in the Section on Experimental Setup.

Complementary to the SAXS measurement, another X-ray area detector (JungFrau detector (*23*)) was used to collect the wide-angle diffraction signal around the (111) and (200) Bragg-peaks, which enabled monitoring of the melting process. Although the optical laser fluence has negligible variations, the monochromatized X-ray FEL pulse has large intensity fluctuations (*24*). As a result, a transmissive intensity monitor consisting of a 50 µm thick Kapton target and a silicon photodiode, which collected the scattering from the Kapton foil, was placed upstream of the sample to measure the X-ray pulse energy on a pulse-to-pulse basis. The measured X-ray pulse energy was then used to normalize both the averaged WAXS and SAXS signals.

Examples of the normalized wide-angle scattering patterns measured at selected delays between the laser and X-ray pulses are displayed in the first row of Fig. 2. The corresponding radially averaged WAXS profile evolution is summarized in Fig. 3 (**B**). At $\delta t$ = -5 ps, the profile features well-defined powder rings corresponding to the (111) and (200) Bragg peaks of Au. The laser pump pulse generates hot electrons, and within a few picoseconds, the excess energy is transferred to the lattice through electron-phonon coupling (*25*). This leads to lattice melting and the emergence of the liquid phase. This process was signified by a decrease in the intensity of the Bragg peaks, and the increase of the diffuse scattering apparent as the elevated background on which the Bragg peaks reside (*26*). After about 20-50 ps, the melting process is complete, as the WAXS does not change at later delay times. We note here that the Bragg peaks do not decay to zero; instead, a residual signal of about 1% remains. This is attributed to the diffraction of extended low-intensity wings of the focused X-ray beam, which probe almost unexcited, 'cold' material exposed to low fluences.

The second and third rows in Fig. 2 show the difference SAXS signal, which was obtained by subtracting the normalized reference pattern measured prior to laser excitation (at a negative delay $\delta t$ = -5 ps) from the normalized SAXS patterns measured at $\delta t \geq 0$. The reference pattern is labeled as -5 ps in the second row of Fig. 2. The small-angle scattering observed in the reference pattern can be mainly attributed to the grain structure of the polycrystalline films (see Supplementary Fig. **S1**). By subtracting this reference pattern, we retrieve the net intensity change, $\Delta S(\mathbf{Q}, \delta t) = S(\mathbf{Q}, \delta t) - S(\mathbf{Q}, -5\text{ ps})$, where **Q** is the scattering vector). This intensity change is used in the analysis of the SAXS signal in the remaining part of this paper.

Immediately after the melting process is complete, the intensity of the SAXS signal starts to increase and gradually evolves into a pattern with several ring-like features observed on the detector tuned to the smallest scattering angles, *i.e.*, the lower right detector in Fig. 1 (**C**). The intensity increase also extends to the other three detectors as well, albeit on a smaller scale. Starting from 75-100 ps, the whole detector lights up, and the ring-like pattern becomes more pronounced. The diffraction pattern rapidly evolves, with the rings shifting to smaller scattering vectors and merging with each other as time progresses from 100 to 500 ps. After 500 ps, the total intensity moves towards the beam stop area, and finally decreases at delays from 750 to 1500 ps.

The rapid rise of the small-angle intensity provides direct evidence for the formation of nanoscale density heterogeneities with characteristic length-scales on the order of tens of nanometers. The rapidly evolving ring pattern suggests a complex nature of the transient laser-generated structures characterized by several dominant length-scales, while the increase of the signal intensity and the evolution of the ring pattern towards smaller scattering angles indicate a continuous growth and coarsening of the nanostructures within the expanding ablation plume. The interpretation of the rich data obtained from the time-resolved SAXS measurements in terms of the actual dynamics of the film disintegration is not straightforward and requires assistance from the atomistic modeling discussed below, in the Section on Simulation Results.

Since the scattering signal is isotropic, we performed an azimuthal average on the net scattering intensity change, denoted here as $\Delta S(Q, \delta t)$. This allows us to summarize the time evolution of the SAXS signal using a two-dimensional (2D), false-color representation displayed in Fig. 3 (**A**). The multiple rings mentioned above are visible as stripes. The stripes are tilted with respect to the vertical direction, which reflects their shift towards smaller Q as time progresses. To quantify the onset time of the small-angle scattering signal, the SAXS Q range is divided into two regions, denoted here as "low Q" (0.005 Å$^{-1}$ - 0.026 Å$^{-1}$) and "high Q" (0.031 Å$^{-1}$ - 0.096 Å$^{-1}$). The time dependencies of the intensity averaged over the two Q regions are shown in Fig. 3 (**C**). In the low Q region, the signal increases gradually with time, while it exhibits an abrupt change at around 50 ps in the high Q region. The simultaneous WAXS measurement is shown in Fig. 3 (**B**), and the changes of the integrated intensity of the Au (111) peak extracted by fitting the signal with a Gaussian on a linear baseline (see Supplementary Material Section **2**) are shown in Fig. 3 (**D**). The black line in Fig. 3 (**D**) represents an exponential fit to the derived integrated peak signal and yields a decay time of (4.4±0.1) ps. Comparing with the two curves plotted in (**C**), it is clear that the completion of the melting process precedes the increase of the SAXS signal.

The fluence dependence of the evolution of the SAXS profiles is illustrated for the 100 nm film by the 2D Q-resolved maps in Fig. 4 (**A-D**) and the Q-averaged plots in Fig. 4 (**E**) for several incident laser fluences. It is noteworthy that the overall intensity of the small angle scattering is the highest at the lower incident fluences of 2.5 and 4.2 J/cm$^2$, Fig. 4 (**A, B**), suggesting the appearance of relatively large density heterogeneities on the timescale of about 100 ps. In comparison, the intensity increase of the SAXS signal for incident fluences of 6.3 and 7.7 J/m$^2$ is smaller, but the SAXS patterns exhibit a modulation of the broad SAXS peak intensity, Fig. 4 (**C**, **D**). The latter observation hints at the emergence of a more complex structure of the ablation plume characterized by multiple length-scales. The temporal evolution of the low/high Q-averaged SAXS signal for the four fluences is summarized in Fig. 4 (**E**). As discussed above, the intensity is the highest at a fluence of 4.2 J/cm$^2$ in both Q-regions, and there is a protracted response of the SAXS onset time at higher laser fluence, *i.e.*, at 7.7 J/cm$^2$. Interestingly, the drop of the intensity with increasing fluence is less pronounced in the higher Q region, where the contribution comes from smaller structural features.

For the 250 nm film, the fluence dependence of the scattering intensity is less pronounced, especially in the low-Q region, as can be seen from Fig. 4 (**F-J**). Nevertheless, the scattering intensity extends to higher Q values with increasing laser fluences, leading to the formation of a broader peak of the scattering intensity. These results point to the formation of transient density heterogeneities with broader size distributions in the course of the ablation process.

The fluence dependence of the WAXS and SAXS signals is further visualized in Fig. 5, where the delay δt is set to 250 ps for the 100 nm film and 200 ps for the 250 nm film, after completion of the melting process and the appearance of the SAXS signal in all the experiments discussed above, and the incident laser fluence $F_{inc}$ is scanned in a wide range, from 0 to 11 J/cm$^2$.

As can be seen from Fig. 5 (**C**, **D**) for the 100 nm film, the difference SAXS signal appears only when the film is completely molten, *i.e.*, above 1.1 J/cm$^2$. Below the fluence threshold for complete melting, the film expands due to laser heating. This is the reason why the (111) and (200) peaks in Fig. 5 (**B**) shift towards smaller Q, and this shift becomes more pronounced as the laser fluence increases. Above the threshold for complete melting, the SAXS intensity increases rapidly, reaching the maximum at around 3 J/cm$^2$ for the low Q region and at around 4 J/cm$^2$ for the high Q region. Fig. 5 (**A**) gives more information regarding the Q dependent nanoscale ablation dynamics: with increasing laser fluence, the single peak broadens, shifts to higher Q values, and subsequently develops a structure featuring several low-intensity modulations or sub-peaks. The SAXS intensity eventually diminishes (faster at low Q), indicative of a progressive homogenization of the ablation plume.

For the 250 nm film, the increase of the SAXS signal starts at about 1.7 J/cm$^2$, while complete melting occurs only above 3 J/cm$^2$, indicating that the ablation starts even before the film melts through its full thickness. Since the film thickness is larger than the commonly accepted electron ballistic transport length in Au, the excitation of the material is spatially inhomogeneous over the depth of the film. Therefore, ablation can take place in a near-surface region, where the deposited energy density is highest, while the deeper regions of the film are still solid. Correspondingly, the threshold fluence for the ablation onset at the surface is smaller than that required for the complete melting of the film. Compared to the data for the 100 nm film (Fig. 5 (**A**) the SAXS-pattern of the 250 nm film exhibits a broader peak without any pronounced sub-structure. We also attribute this to the non-uniform energy distribution throughout the thickness of the film, where the depth dependence of the material decomposition dynamics leads to the emergence of density inhomogeneities over a continuous range of length scales. The superposition of scattering contributions from different parts of the film can then produce a broad featureless SAXS peak observed for the thicker film.

In the following sections, we use the results of large-scale molecular dynamics simulations to establish a link between the dynamics of material decomposition in laser ablation and the characteristics of the experimental SAXS profiles. In particular, we demonstrate how the spatially heterogeneous phase decomposition manifests itself in the evolution of the SAXS patterns. This approach enables us to identify the SAXS signatures of the formation and growth of nanovoids, the transient appearance of molten filaments and the ejection of nanodroplets in different ablation regimes, such as photo-mechanical spallation and phase explosion.

**Simulation results**

The links between the dynamics of phase decomposition in laser ablation and its manifestation in the evolution of SAXS patterns are established in this study by performing large-scale atomistic modeling of laser ablation of Au films and calculating the SAXS patterns for the atomic configurations predicted in the simulations. The computational setup is chosen to represent the systems studied in the pump-probe experiments described above. Due to the high computational cost of the atomistic simulations, we restrict ourselves to the case of a 100 nm Au film. The film is deposited on a 100 nm thick $Si_3N_4$ substrate and irradiated with a 50 fs laser pulse at a wavelength of 800 nm. The lateral size of the computational system is $200 \times 200$ nm, and the total number of Au atoms is 235,298,000.

The simulations are performed with two values of the absorbed fluence $F_{abs}$, 0.1069 J/cm$^2$ and 0.2014 J/cm$^2$. These two fluences are in the range where the film disintegration is characterized by the appearance of nanoscale density heterogeneities, which can be probed in SAXS measurements. The lower fluence is in the regime of photomechanical spallation, while the higher fluence corresponds to the transition from spallation to phase explosion. Outside this fluence range, the laser-induced processes, such as melting at lower fluences and an explosive phase decomposition into a mixture of vapor and small nanodroplets at higher fluences, do not involve the formation of structural features that contribute to the small angle scattering, thus making the results of SAXS probing less informative.

As a reference, using experimental properties of Au (*27,28*), the energy density required for heating the film to $T_m = 1337$ K and fully melting it at $T_m$ is 41.8 kJ/mol (corresponding to an absorbed fluence of 0.041 J/cm$^2$ for a 100 nm film), while heating the film to the thermodynamic critical temperature $T_c = 7400$ K (*28*) requires the deposition of 241.7 kJ/mol or 0.24 J/cm$^2$ (see Supplementary Material Section **5**). The corresponding values calculated with the thermodynamic properties predicted by the Embedded Atom Method (EAM) interatomic potential (*29*) used in the simulations are 42.3 kJ/mol or 0.0415 J/cm$^2$ for the complete melting of the film at $T_m^{EAM}$ = 1318 K and 328.9 kJ/mol or 0.323 J/cm² for heating the film to $T_c^{EAM}$ = 9250 K (see Supplementary Material Section **6**). Thus, the deposited energy density in the lower-fluence simulation exceeds the threshold for complete melting by a factor of 2.6, while the deposited energy density in the higher-fluence simulation is 1.6 times lower than the energy needed for homogeneously heating the film to $T_c$.

Snapshots of atomic configurations from the lower-fluence simulation are provided in Fig. 6 (**A**), while the evolution of pressure, density, and temperature is shown in the form of contour plots in Fig. 7 (**A, C, E**). Following laser excitation at the start of the simulation, the rapid electronic heat transfer and the electron-phonon coupling determine the initial energy redistribution and lattice heating, Fig. 7 (**E**). The temperature rises above the limit of lattice stability against melting, ~$1.25 T_m$ (*30*), throughout the thickness of the film, and the film fully melts by 14 ps. The rapid heating occurring under stress confinement conditions (*10-12*), *i.e.*, faster than the film expansion can occur, leads to the generation of high compressive stresses (up to a maximum level of 28 GPa detected at a depth of 56 nm at 14 ps) in the interior of the film, Fig. 7 (**A**). The subsequent ejection and disintegration of the film is largely driven by the dynamics of the relaxation of these initial compressive stresses. Here and below, the stress level is characterized by the pressure, which is the negative one third of the first invariant of the stress tensor.

The stress relaxation in the presence of a free surface leads to the generation of a bipolar pressure wave consisting of a compressive component followed by a tensile one, Fig. 7 (**A**). When the compressive component of the wave reaches the substrate, it is partially reflected from the gold-substrate interface, with the nature of the reflection defined by the acoustic mismatch between the two materials. Using the values of the elastic moduli and densities of Au (*29*) and $Si_3N_4$ (*31, 32*), the values of the specific acoustic impedance of the two materials can be estimated as $Z_{Au}$ = 64×10⁶ Pa s/m and $Z_{Si_3N_4}$ = 25×10⁶ Pa s/m. For a pressure wave coming to the Au-$Si_3N_4$ interface from the Au side, the reflection and transmission coefficients for stress amplitudes can be calculated as $R_{Au \to Si_3N_4} = (Z_{Si_3N_4} - Z_{Au})/(Z_{Au} + Z_{Si_3N_4}) = -0.44$ and $T_{Au \to Si_3N_4} = 2Z_{Si_3N_4}/(Z_{Au} + Z_{Si_3N_4}) = 0.56$.

The partial transmission of the compressive wave to the substrate layer ($T_{Au \to Si_3N_4}$ = 0.56) accelerates the layer away from the Au film, while the interfacial spallation prevents the layer from being pulled back by the tensile component of the bipolar wave. As a result, the $Si_3N_4$ layer acquires a momentum in the negative direction and moves away from the Au film with a constant velocity of ~612 m/s. The transmission of the pressure wave into the substrate layer triggers long-term elastic oscillations of the layer, with the maximum tension reaching -8 GPa. The $Si_3N_4$ layer is "unbreakable" in its coarse-grained representation used in the model, but can be expected to disintegrate in the experiments.

Due to the negative sign of $R_{Au \to Si_3N_4}$, the compressive component of the stress wave is partially reflected from the interface as a rarefaction wave. This reflected wave combines with the stronger tensile wave propagating from the surface, leading to the concentration of tensile stresses inside the Au film and at the gold-substrate interface. The tension at the interface is sufficient for the separation/spallation of the Au film from the substrate layer, which can be seen to occur at ~50 ps from the density contour plot in Fig.7 (**C**).

Inside the Au film, the tensile stress concentration is observed at 40-50 ps in a broad region outlined by the -3 GPa isobar in Fig. 7 (**A**). The highest tensile stress reaches -5.6 GPa in the lower part of the film at 48 ps. The combination of heating and tension results in a massive nucleation of multiple voids throughout a broad central region of the film. As discussed in (*33*), the region affected by the void nucleation can be described within the framework of classical nucleation theory, as a region where the free energy barrier for the void nucleation drops below a certain threshold value. Following the nucleation at ~50 ps, the voids grow and coalesce, leading to the formation of a continuous foamy structure of interconnected liquid regions, which can be seen in the snapshot shown in Fig. 6 (**A**) for 100 ps.

As the Au film continues to expand, two distinct regions emerge from the evolving foamy structure. The expansion of the upper and hotter part of the foamy structure is faster, and the liquid regions are gradually stretched into filaments elongated in the direction of the film expansion. The expansion of the lower part is slower, and the liquid structure evolves into an arrangement of large voids sandwiched between two liquid layers. The upper of these two layers, marked by the black arrow in the snapshots shown for 500 ps in Fig. 6 (**A**), separates the two distinct regions of the foamy structure. The upper part features a more fine-grained structure of liquid filaments extending to the upper liquid layer, while the lower part has a coarser structure of large voids. While the film ejection and disintegration produce a pronounced density variation along the direction normal to the film surface, this variation cannot be probed by an X-ray beam pointed in the same direction. However, the SAXS patterns should be sensitive to the emergence of the two regions of the transient foamy structure featuring distinct length scales of the density variation in the lateral (parallel to the surface) directions. The analysis of the connections between the computational predictions and the experimental observations provided in the next section will therefore focus on the lateral density variations emerging in different parts of the film during the spallation process.

The results of the higher-fluence simulation, performed at almost twice the energy density deposited by the laser pulse, are illustrated by snapshots of atomic configurations shown in Fig. 6 (**B**) as well as the contour plots of pressure, density, and temperature shown in Fig.7 (**B**, **D**, **F**). While the deposited energy is 1.6 times lower than that required to uniformly heat the film to $T_c$, the maximum temperature at the film surface exposed to the laser pulse reaches $0.947T_c$ at about 15 ps. The top layer of the film rapidly heated up to and above ~$0.85T_c$ undergoes a rapid decomposition into vapor and liquid droplets in a process commonly referred to as phase explosion (*4, 10, 34*). The recoil pressure generated by the phase explosion at the top of the film pushes the underlying material down and is responsible for the formation of a thin liquid layer marked by the black arrow in a snapshot shown for 500 ps in Fig. 6 (**B**).

As can be seen from the snapshots, this liquid layer separates the upper region directly affected by the explosive phase decomposition from a larger underlying region where the formation and uniaxial expansion of a foamy structure is observed. While the evolution of the foamy structure and the formation of vertically aligned liquid filaments has some similarity to those discussed above for the lower-fluence simulation, there are some notable differences. In particular, the filaments have a much finer scale and are surrounded by vapor phase atoms represented in the snapshots in Fig. 6 (**B**) as red dots. The presence of a substantial amount of vapor throughout the thickness of the expanding film indicates the transition to a film disintegration regime, where both the relaxation of laser-induced stresses and the rapid release of the vapor phase contribute. Indeed, the level of tensile stresses supported by a part of the film outlined by the $0.7T_c$ isotherm in Fig. 7 (**B**) does not exceed -0.8 GPa, and the pressure remains positive at all times in the upper region undergoing the phase explosion. The tensile stresses are still generated in the lower and colder part of the film, although the stress magnitude decreases significantly with respect to the lower fluence simulation, Fig. 7 (**A**), due to a combined effect of the ablation recoil generated by the phase explosion and the reduced ability of the hot molten material to support the dynamic tensile loading.

The observation of the foamy nanostructure evolving into an interconnected network of elongated liquid filaments is of direct relevance for the interpretation of the data obtained by SAXS probing of the ablation dynamics. Similar to the lower-fluence simulation, the transient nanostructures generated in the course of the phase decomposition exhibit a range of characteristic length scales of the density variation in the lateral directions. The mapping of the transient nanoscale structures predicted in the simulations to the SAXS patterns measured in the experiments is discussed next.

## Discussion

The atomistic simulations described above provide detailed information on the dynamics of the film decomposition and ejection in the regime of photomechanical spallation ($F_{abs} = 0.1069$ J/cm$^2$) and at the transition to the phase explosion regime ($F_{abs} = 0.2014$ J/cm$^2$). The conversion of the absorbed fluence $F_{abs}$ to the incident fluence $F_{inc}$ used in the experiments is not straightforward due to the uncertainty in the fluence dependence of the effective (integrated over the duration of the laser pulse) reflectance of the Au films. While the intrinsic normal incidence reflectance calculated for Au at a wavelength of 800 nm using Fresnel's equations is 0.974 (*35*), the effective surface reflectance under experimental conditions can be altered by surface roughness and the electronic excitation during the laser pulse. The experiments are performed for polycrystalline films, as evidenced by the two powder diffraction rings in the WAXS-signal of the unpumped sample in Fig. 1 (**A**). The SEM and AFM images in the Supplementary Fig. S1 show that the average grain size is about 50 nm and the peak-to-peak surface roughness is about 8 nm. This surface roughness results in a small reduction of the surface reflectance down to 0.968, as predicted by electromagnetic calculations performed with a model described in Section **4** of the Supplementary Material for computational samples reproducing the surface morphology of the experimental 100 nm films. An additional and stronger reduction of the reflectance is expected from the transient increase in the electron temperature and the associated changes in the electron scattering rates during the laser pulse. However, this reduction is difficult to quantify theoretically, as it has been demonstrated that the experimental data for Au cannot be adequately described by a Drude-Lorentz model without the introduction of empirical *ad hoc* corrections (*36*). Indeed, the direct application of the Drude model converts the absorbed fluences used in the simulations to incident fluences of 1.0 J/cm$^2$ and 1.4 J/cm$^2$, which is comparable to the experimental threshold for melting of the film, about 1.1 J/cm$^2$.

Given the uncertainty of the theoretical conversion between $F_{inc}$ and $F_{abs}$, we resort to an approximate matching of the fluence ranges using the thresholds for film melting and disintegration. Assuming that the deposited energy at the incident fluence threshold for complete melting of the 100 nm film (1.1 J/cm$^2$, as determined from WAXS probing) corresponds to the deposited energy density needed for the complete melting of the film (0.041 J/cm$^2$, as estimated in the previous section), the effective reflectance is about 0.963. This value is higher than that predicted theoretically and measured in previous experiments at this fluence (*36*), suggesting either an uncertainty in the experimental fluence calibration or the existence of energy loss channels activated during and immediately after laser excitation. According to the simulations, the energy transfer to the $Si_3N_4$ substrate (energy of the center-of-mass motion of the substrate away from the film and energy of the elastic wave generated in the substrate, Fig. 7 (**A, B**)) accounts for 0.007 and 0.017 J/cm$^2$ at $F_{abs}$ of 0.1069 and 0.2014 J/cm$^2$, respectively. The energy loss due to electron emission has been estimated to be negligible at fluences near the melting threshold (*30*) and, in the absence of other obvious channels of energy loss, we have to assume that the effective reflectance remains surprisingly high up to the melting threshold of the 100 nm film. As the incident fluence increases above the melting threshold, a substantial drop of the effective reflectance is expected, leading to a highly nonlinear scaling of the absorbed fluence with the incident one. Based on these considerations, we approximately match the conditions of the low-fluence ($F_{abs}$ = 0.1069 J/cm$^2$) and high-fluence ($F_{abs}$ = 0.2014 J/cm$^2$) simulations with the experimental data measured at $F_{inc}$ of 4.2 J/cm$^2$ and 6.3 J/cm$^2$, respectively.

The X-ray scattering profiles calculated for the atomic configurations predicted in the simulations for different delays after the laser excitation are compared with the corresponding profiles obtained in the experimental measurements in Fig. 8 (**A, B** and **D, E**). The computational and experimental SAXS profiles exhibit similar shapes and shape evolution with time. The appearance of broad small-angle scattering peaks gradually shifting towards lower Q values is observed in all cases. To quantify the time evolution of the scattering profiles, the Q-averaged intensity evolution is calculated for the low- and high-Q regions defined above, in the discussion of the experimental results, Fig. 8 (**C, F**). The trends show a remarkable quantitative agreement between the computational predictions and the experimental observations. The striking similarities include the delayed onset times for the increase of the low-Q region intensity and the drop of the high-Q intensity at delays larger than about 250 ps. The increase in fluence leads to a slightly delayed increase in the low-Q intensity and the formation of a broader peak at low Q values at larger delays.

The good agreement between the computational and experimental SAXS profiles gives us confidence in the realistic representation of the film disintegration dynamics in the simulations and encourages a more detailed analysis of the specific transient features of the film disintegration responsible for the evolution of the SAXS profiles. As noted in the discussion of the simulation results, at both fluences, the phase decomposition proceeds differently in parts of the expanding film located at different depths below the film surface. While the SAXS probing in normal-incidence transmission geometry contains information averaged over the depth of the film, analysis of the computational results can help to disentangle the contributions from different depths, where the phase decomposition produces nanostructures with distinct length scales of the density variation in the lateral directions. To resolve the depth-dependent contributions to the SAXS profiles, we first identify regions of distinct evolution of the transient nanoscale structures and then perform calculations of the "partial" SAXS profiles contributed by these regions.

In the low-fluence simulation ($F_{abs}$ = 0.1069 J/cm²), the two regions of distinct nanostructure are separated by a molten layer marked by the black arrow in Fig. 6 (**A**). The 2D real-space projected density maps for these two regions for a time delay of $\delta t$ = 200 ps are shown in Fig. 9 (**A2** and **A3**), while the total density map combining the two contributions is shown in Fig. 9 (**A4**). Despite the clear visual difference between the regions in a snapshot shown for 200 ps in Fig. 6 (**A**), the projected density distributions in Fig. 9 (**A2** and **A3**) are quite similar. Apparently, the 2D projections of the larger vertically elongated voids in the upper region of the expanding film and smaller closer-to-spherical voids in the lower region yield similar density distributions when projected on the plane of the film. As a result, the partial contributions to the SAXS profile from the upper and lower regions, shown by the dotted and dashed lines in Fig. 9 (**A1**), have similar shapes. When combined, they yield a single peak shown in blue in Fig. 9 (**A1**). The peak is centered at about $Q = 0.021$ Å$^{-1}$ (marked by the blue arrow in Fig. 9 (**A1**), which corresponds to a length-scale of $d = \frac{2\pi}{Q} \approx 30$ nm (shown by the length of the magenta segment in Fig. 9 (**A4**)).

At $\delta t$ = 500 ps, both regions expand and coarsen (Fig. 6 (**A**)), which is reflected in the 2D density maps shown in Fig. 9 (**A5** and **A6**). The structure is somewhat coarser in the lower region, which is also reflected by the relative positions of the corresponding peaks of the partial SAXS profile. Namely, the peak produced by the lower region, shown by the dashed orange line in Fig. 9 (**A1**), is shifted to the lower-Q direction with respect to that produced by the upper region. This shift, however, is relatively small, and the two partial profiles merge into a single broader peak of the total SAXS profile shown by light orange color in Fig. 9 (**A1**). The peak position marked by the orange arrow in Fig. 9 (**A1**) is $Q = 0.0091$ Å$^{-1}$, which corresponds to the characteristic length-scale of $d \approx 69$ nm, shown by the length of the magenta segment in Fig. 9 (**A7**). Therefore, the computational prediction for the SAXS signature of ablation in the spallation regime is the appearance of a relatively narrow single peak shifting in the direction of smaller Q with time. This behavior is consistent with the experimental data obtained at laser fluences close to the threshold for film disintegration, Fig. 8 (**B**), although the Q values of the peak positions are smaller in the experiment (Q = 0.012 Å$^{-1}$ and Q = 0.008 Å$^{-1}$ at $\delta t = 200$ ps and $\delta t = 500$ ps, respectively). The shift of the SAXS peaks in the lower Q direction in the experiments can be traced back to the early appearance of strong low-Q signal as early as 100 ps after the laser excitation, Fig. 8 (**B**), which is absent in the simulations, Fig. 8 (**A**). This effect is common for the lower and higher fluence conditions, and is briefly discussed below.

In the high fluence simulation ($F_{abs} = 0.2014$ J/cm$^2$), the two regions that can be distinguished in the snapshots shown in Fig. 6 (**B**) are the region above the molten layer marked by the black arrow in the snapshot shown for 500 ps and the lower region located between the two thin liquid layers. The upper region is generated by an explosive release of vapor and consists of nanodroplets and small liquid filaments attached to the liquid layer. The lower region, generated by a combined effect of mechanical stretching and vaporization, develops into a foamy structure of vertically elongated interconnected liquid filaments. The difference in the dynamics of the phase decomposition in these two regions is more prominent as compared to the lower-fluence simulation, which is reflected in the corresponding 2D density maps shown in Fig. 9 (**B2** and **B3**) for a delay of 200 ps and Fig. 9 (**B5** and **B6**) for 500 ps. Already at $\delta t = 200$ ps, the top layer undergoing the explosive phase decomposition features a finer-grained density distribution, as reflected by the partial SAXS profile shown by the dotted blue line in Fig. 9 (**B1**). Due to the small mass fraction of this layer, however, it makes only a minor contribution to the total SAXS profile, which is dominated by the large lower region of the expanding film. The total SAXS profile is centered at Q = 0.024 Å$^{-1}$, which corresponds to the characteristic length-scale of $d = 26$ nm, shown by the length of the magenta segment in Fig. 9 (**A4**).

The difference in the characteristic length-scales of the density distributions in the upper and lower layers further increases with time, as can be seen from the 2D density maps for $\delta t = 500$ ps shown in Fig. 9 (**B5** and **B6**). In this case, despite the low mass fraction of the upper layer, its contribution turns the shoulder of a broad partial SAXS profile of the dominant lower layer into the second peak in the total SAXS profile. The two peaks of the total SAXS profile, marked by two orange arrows in Fig. 9 (**B1**), are located at Q values of 0.0147 Å$^{-1}$ and 0.008 Å$^{-1}$, which corresponds to the characteristic length-scales $d$ of 43 nm and 79 nm, respectively. The smaller length-scale and the corresponding peak in the SAXS signal has a substantial contribution from the upper layer.

The broadening of the SAXS peak through the expansion into lower Q direction with increasing laser fluence is consistent with experimental observations, Fig. 5 (**A**). According to the simulations, this broadening can be attributed to the transition from spallation to the phase explosion regime of film disintegration, with the latter producing a mixture of vapor and small liquid droplets that contribute to the scattering at larger Q. Indeed, the overall drop in the SAXS signal at the maximum fluences probed in the experiments, Fig. 5 (**A**), may indicate the completion of the transition to the phase explosion regime of film disintegration throughout the film thickness.

The splitting of the SAXS profile into sub-peaks in the simulated SAXS profiles, Figs. 8 (**D**) and 9 (**B1**), is a consequence of the coexistence of different regimes of material decomposition activated at different depth below the irradiated surface. Again, this is (at least qualitatively) consistent with the experimental observations, *i.e.*, the appearance of rings in the 2D SAXS patterns in Fig. 2, the corresponding stripes in Fig. 3 (**A**), and sub-peaks in Fig. 8 (**E**). However, the quantitative difference in the Q-positions and the larger magnitude of the sub-peaks in the experimental profiles, as well as their persistent presence in a broad range of fluences (Fig. 5 (**A**)), indicate that additional processes/effects not considered in the simulations contribute to the experimental SAXS patterns.

One of the effects not accounted for in the simulations is the nanoscale surface roughness of the polycrystalline Au film. This roughness may lead to a spatial variation of the energy deposition due to plasmonic effects (*37*, *38*), while the results of large-scale atomistic modeling (*39*) suggest that the spatial modulation of the energy deposition can lead to a modulation of density in the ablation plume persevering for hundreds of picoseconds. As illustrated by the Supplementary Fig. S3, the electromagnetic simulation of the laser interaction with a model system reproducing the surface roughness of an experimental film reveals an uneven energy deposition leading to the formation of "hot" and "cold" spots in the electron and lattice temperature distributions. The analysis of the implications of the energy deposition on the dynamics of phase decomposition and its SAXS signature is the subject of our current work, which will be reported elsewhere.

We also note that, due to the small lateral size of the computational cell (200 nm), processes occurring on spatial scales beyond 100 nm (Q $\leq$ 0.006 Å$^{-1}$) cannot be reliably described. On the other hand, it is well known that so-called Laser-Induced Periodic Surface Structures (LIPSS) with feature sizes ranging from approximately 0.1 to 1 µm represent a common phenomenon on laser-irradiated surfaces at the excitation levels addressed in this study, *e.g.*, see (*40*) and references therein. Since, in the ablative regime, LIPSS form on a timescale of tens of picoseconds to nanoseconds (*39*, *41*), it is very likely that such processes contribute to the experimental SAXS patterns, in particular at low Q-values. Indeed, the rapid rise of the low-Q SAXS signal observed in Fig. 8 (**B**, **E**) already at $\delta t$ = 100 ps represents the most striking difference with the computational profiles shown in Fig. 8 (**A**, **D**). The contribution of processes responsible for the rapid emergence of the low-Q signal can be expected to persist at longer delays, leading to the quantitative differences between the experimental and simulated SAXS profiles.

In summary, in this work we demonstrate that the combination of X-ray probing and large-scale atomistic modeling of ultrashort pulse laser ablation of thin Au films represents a powerful approach to characterize the rapid spatially heterogeneous nanoscale phase decomposition driven by a combination of extreme photomechanical and thermodynamic driving forces. In agreement with the simulations, the WAXS measurements show that ablation and phase decomposition are preceded by melting and occur from the liquid state. Furthermore, they allow to establish the threshold for the loss of crystalline order in the film, thus enabling the alignment of the scales of energy deposition in modeling and experiments. The direct comparison of the SAXS profiles predicted for the simulated atomic configurations with the measured ones enables the identification of the characteristic SAXS signatures for the photomechanical spallation and phase explosion regimes of laser ablation, thus confirming the predicted coexistence of different processes in different parts of the irradiated films.

The insights into the origins of the SAXS profiles provided in this work open a new avenue for the investigation of rapid phase transformations in materials subjected to ultrafast energy deposition and brought to extreme states of thermal and mechanical nonequilibrium. Future work will include X-ray probing at non-normal incidence to access the emergence of density variations in both the lateral and axial directions at the initial stage of the ablation process.

## Materials and Methods

### Experimental setup

The experiment was performed at the X-ray Pump Probe instrument of the Linac Coherent Light Source (*42*). A schematic of the experimental setup is shown in Fig. 1 (**A**). The 800 nm optical laser and 9.5 keV X-ray pulse incident on the sample were close to colinear. The focal spot size of the laser is $\sigma_l \approx 28$ μm (RMS) (see Supplementary Material Section **7**), and the pulse duration is estimated to be 50 fs. The X-ray FEL pulses were monochromatized by a double reflection Diamond (1,1,1) monochromator (*24*) and focused on the sample using beryllium compound refractive lenses with a focal length of 1.5 m and a focus of 1.2 μm (RMS). Since the laser spot size is much larger than that of the X-ray beam, we assume that the X-ray probed region was uniformly irradiated. The laser profile is close to Gaussian, which allows us to estimate the peak fluence of the beam to $F_{inc} = J/(2\pi\sigma_l^2)$ given the measured laser pulse energy J. $F_{inc}$ is referred to as the incident laser fluence throughout the paper. The samples are 100 and 250 nm thick Au films prepared by a process known as anodic vacuum arc deposition (*43*). They are deposited on arrays (120 by 120) of 100 nm thick $Si_3N_4$ membrane windows, which are manufactured from 4-inch silicon wafers. The wafer is mounted on translation stages perpendicular to the incoming X-ray beam, with in-plane motions synchronized with the laser and X-ray pulse delivery. As a result, each laser and X-ray pulse pumps and probes a new window consisting of a fresh sample. The sample assembly was placed in a vacuum chamber. This sample scheme has been used in several single pulse type experiment performed at X-ray FELs to study irreversible processes initiated by an optical laser (*44*). A high-resolution microscope and a wide field of view camera were implemented to monitor the sample, the laser, the X-ray interaction point to check the spatial overlap, the ablation imprint as well as to navigate through the sample on-the-fly. During the experiment, we adjusted the laser fluences and X-ray laser delays while recording the sample X-ray scattering using a JungFrau detector (*23*) and four ePix100 detectors (*22*). They were placed 0.15 meter and 5.05 meter downstream of the sample to simultaneously follow the evolution from Bragg peaks to liquid rings at wide angles and the formation of nanoscale structures at small angles. The grainy features in the SAXS pattern are known as speckles (*45*), a result from the incident X-ray FEL pulses being nearly transversely coherent.

**Computational model**

The laser ablation simulations of an Au film deposited on a $Si_3N_4$ membrane are performed with a hybrid atomistic-continuum computational model combining the solution of electromagnetic wave equations (EM) to simulate the laser energy deposition on the irradiated film, the two-temperature model (TTM) to represent the electron-phonon equilibration and energy redistribution by the electron thermal conductivity, a classical atomistic molecular dynamics (MD) model to simulate laser-induced nonequilibrium phase transformations and Au film disintegration dynamics, and a one-dimensional (1D) coarse-grained (CG) MD representation of the mechanical interaction of the Au film with the $Si_3N_4$ substrate.

The combined TTM-MD model for the simulation of laser interactions with metals is well described in the literature (*14*,*30*,*46*), and here we only provide the model parameters used in the simulations reported in the present paper. The initial 100 nm Au film is represented by a single crystal slab with a face-centered cubic (fcc) structure and (001) orientation of the free surface. The slab has a thickness of 100 nm, lateral dimensions of 200 nm by 200 nm, and consists of 235,298,000 Au atoms. Periodic boundary conditions are applied in the lateral directions, parallel to the surface of a target. The interatomic interactions between Au atoms are described by an embedded atom method (EAM) potential parametrized with a particular attention to high-temperature material properties relevant to the simulations of laser ablation (*29*). The TTM equation for the electron temperature is solved by a finite difference (FD) method on a three-dimensional (3D) grid with a FD cell size of 1 nm. The FD cells are mapped to the corresponding regions of the MD system, and the local electron temperature enters a coupling term added to the MD equations of motion to account for the energy exchange between the electrons and the lattice (*46*).

The CG MD representation of the 100-nm-thick $Si_3N_4$ membrane window is based on dividing the 100 nm membrane into 1000 layers and representing each layer by a bead capable of 1D motion. The mass of each bead is equal to the mass of the corresponding $Si_3N_4$ layer (assuming the density of 3.19 g/cm$^3$), and the individual beads interact with the neighboring beads through a harmonic potential (Hook's law) fitted to reproduce the elastic modulus of the $Si_3N_4$ film, E = 192 GPa (31). The interaction of Au atoms with the membrane is described by a Morse potential, which is defined as a function of the distance between a metal atom and the plane located at the position of the upper bead in the CG MD representation of the membrane. The parameterization of the Morse potential assumes a work of adhesion between the Au film and $Si_3N_4$ surface of 1.0 J/m$^2$, which is a typical interfacial energy between a metal and a membrane (*47*), as well as the local stiffness of $Si_3N_4$ (31) under uniaxial compression. The resulting equilibrium bond length, potential "width", and energy parameters are $r_e$ = 2.94 Å, $a$ = 1.60 Å$^{-1}$, and $D$ = 550 meV.

The laser energy deposition on the Au film is simulated with a 1D TTM-EM model based on the solution of Maxwell's equations, as briefly described in the Supplementary Material Section **3**. In the TTM-EM model, the permittivity of the Au film is approximated by the Drude model, incorporating a lattice and electron temperature-dependent electron relaxation time. The model captures, in an approximate manner, the enhancement of electron scattering at high electron and lattice temperatures, leading to a nonlinear variation of the optical properties. Within the coupled TTM-EM model, the TTM accounts for the electron heat transport in the Au film during the laser energy deposition and provides the electron and lattice temperature fields for the electromagnetic wave calculation. The results of the TTM-EM calculations are then used as a source term in the TTM-MD model. The TTM-EM calculations are also used to evaluate the effect of surface roughness measured for the polycrystalline films used in the experiments on the surface reflectance and the lateral variation of the laser energy deposition, as discussed in the Supplementary Material Section **4**.

**SAXS profile calculation**

The SAXS profile calculations are performed for transient atomic configurations generated in the atomistic simulations. The atomic configurations are binned to generate 3D number density maps of the Au distribution within the sample system denoted as $\rho(\mathbf{r})$. The maps are generated with a binning pixel size of 2 nm. This size corresponds to $Q \approx 0.314\,\text{Å}^{-1}$, which is larger than the wavevectors considered in the analysis of the SAXS profiles. The total scattering intensity from the sample at a wavevector $\mathbf{Q}$ can be approximated as

$$S(\mathbf{Q}) \approx \left| \int_V f_{\text{Au}}(\mathbf{Q}) \rho(\mathbf{r}) \exp(i\mathbf{Q}\cdot\mathbf{r}) d\mathbf{r} \right|^2$$

$$\approx |f_{\text{Au}}(0)|^2 \left| \int_0^{x_0} \exp(iQ_x x)\, dx \int_0^{y_0} \exp(iQ_y y)\, dy \int_0^{z_0} \rho(x,y,z) \exp(iQ_z z)\, dz \right|^2$$

Here the $z$ direction is along the incident X-ray beam and $x$, $y$ are the two transverse directions. $V = x_0 y_0 z_0$ is the total volume of the simulation system, with $x_0 = y_0 = 200$ nm and $z_0 = 1800$ nm. $f_{\text{Au}}(\mathbf{Q})$ is the atomic form factor of Au. The form factor is approximately proportional to the atomic number $Z$, which allows us to neglect the contribution from the $Si_3N_4$ membranes. Moreover, in the small angle limit, $f_{\text{Au}}(\mathbf{Q}) \approx f_{\text{Au}}(0) \cong Z$. Since the attenuation length in Au under normal conditions is 4 μm at 9.5 keV, the attenuation effect is also negligible. The scattering intensity is, therefore, proportional to the squared norm of the Fourier transform of the Au atom density distribution.

Due to the periodic boundary conditions in the $x-y$ plane in the simulations, we perform 2D Discrete Fourier Transform (DFT) for each simulation layer along $z$. In this way, only the scattering intensities at $Q_x$ and $Q_y$ values multiples of $2\pi/x_0$ and $2\pi/y_0$ are calculated. The pixelwise $Q_z$ component can be derived from the $Q_x$, $Q_y$ 2D map. Each DFT layer is multiplied by $\exp(iQ_z z)$ before being summed up to calculate $S(\mathbf{Q})$.

## Acknowledgments

We thank M. Jerman for preparation of the Au films used as samples in the experiments. Use of the Linac Coherent Light Source (LCLS), SLAC National Accelerator Laboratory, is supported by the U.S. Department of Energy, Office of Science, Office of Basic Energy Sciences under Contract No. DE-AC02-76SF00515. TJA and KST acknowledge support by the Deutsche Forschungsgemeinschaft (DFG, German Research Foundation) through Project No. 278162697-SFB 1242. LVZ, CC, MA, and AV are supported by the National Science Foundation (NSF) through grants CMMI-2302577 and CBET-2126785. LVZ also acknowledges the Research Award from the Alexander von Humboldt Foundation. Computational support was provided by the NSF through the Advanced Cyberinfrastructure Coordination Ecosystem: Services & Support (ACCESS) project DMR110090.

**Author contributions:** YS, DZ and KST conceived the project and designed the experiment. YS, TJA, HL, YC, MD, JMG, MH, MJH, MM, QLN, TS, SS, PS, MS, ST, NW, DZ, KST performed the experiment. YS analyzed the experimental data. CC and LVZ designed the computational model. CC implemented the model and performed the large-scale atomistic simulations. YS, CC, TJA, HL, MA, AV, DZ, LVZ and KST established the comparisons between the experiment and the simulation. YS, CC, LVZ and KST wrote the paper with input from all the other authors.

**Competing interests:** The authors declare no conflicts of interest.

**Data and materials availability:** All data needed to evaluate the conclusions in the paper are present in the paper and/or the Supplementary Materials.

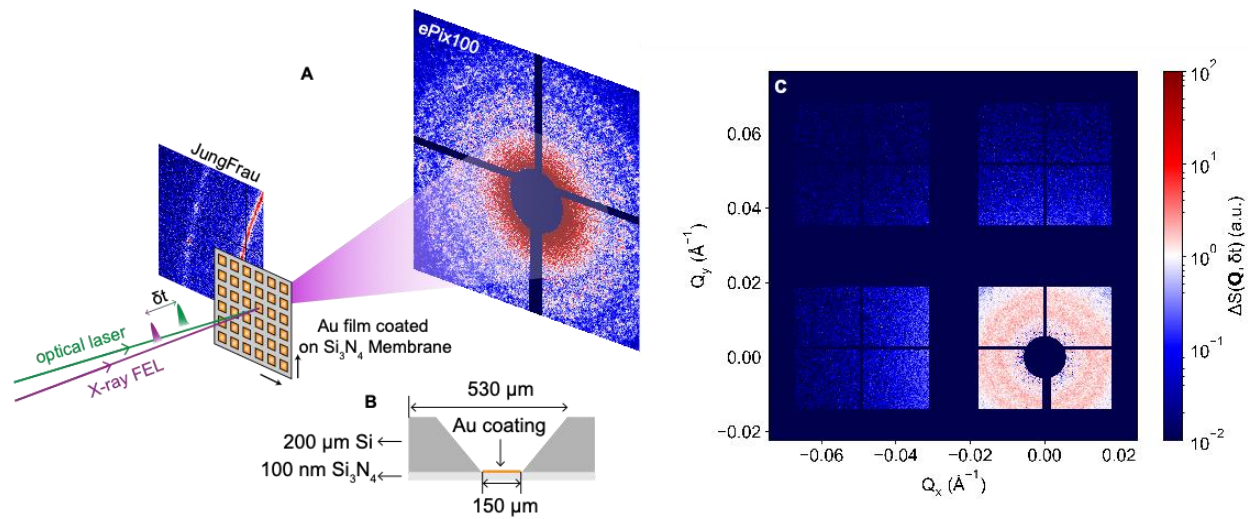

**Fig. 1. Experimental setup with simultaneous small- and wide-angle scattering measurements.** (**A**) Schematic of the femtosecond laser pump, X-ray FEL probe experiment. The samples are 100 nm thick Au films deposited on $Si_3N_4$ membranes made from 4-inch silicon wafers. A zoomed in view of the cross-section of a membrane window is shown in (**B**). The wafer is mounted on translation stages for 2D rastering. The SAXS scattering pattern shown in the schematic is detected by one of the four ePix100s placed at small angles. A JungFrau-1M detector is also used to capture the WAXS from the sample. At negative delays ($\delta t < 0$ ps), it captures Au (111) and Au (200) polycrystalline Bragg peaks from the film as shown in the schematic. (**C**) The difference scattering pattern measured by all four ePix100 detectors. The small-angle images in (**A**) and (**C**) are measured at 250 ps after laser excitation at an incident laser fluence of 6.3 J/cm$^2$.

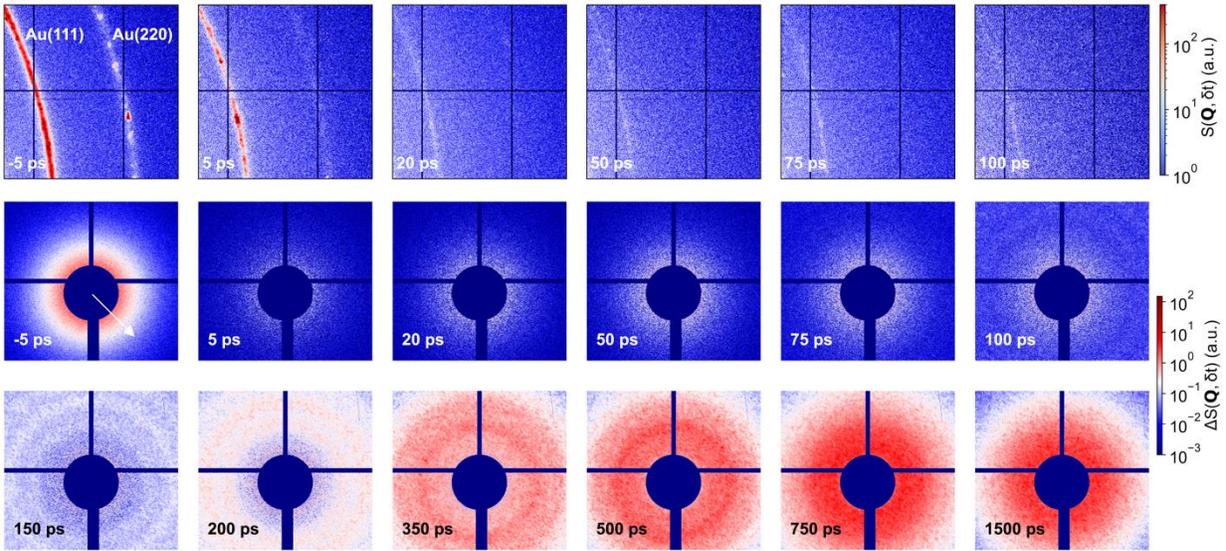

**Fig. 2. Temporal evolution of the WAXS and SAXS patterns measured for a 100 nm film irradiated at an incident laser fluence of 6.3 J/cm$^2$. First Row**: Time resolved WAXS signal. **Second and third row:** SAXS intensity measurement on the detector at the smallest angle. The pattern labeled as -5 ps shows the S(**Q**) pattern used as the reference in the evaluation of the difference patterns ΔS(**Q**, δt)) = S(**Q**, δt) – S(**Q**, -5 ps) plotted in all other SAXS images. The arrow corresponds to a wavevector of 0.01 Å$^{-1}$.

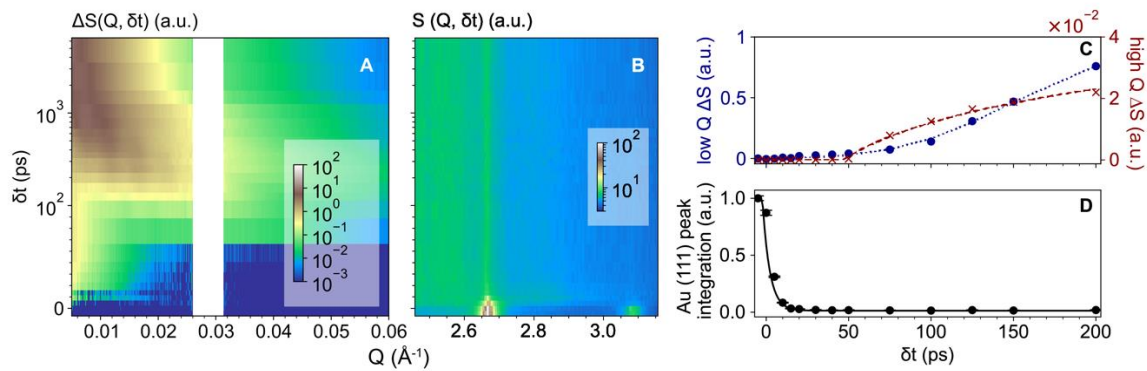

**Fig. 3. Measurement on 100 nm Au film.** Time-resolved radially averaged difference SAXS signal (**A**) and WAXS signal (**B**) obtained at an incident laser fluence of 6.3 J/cm$^2$. The two plots share the same *y*-axis. The white vertical column in (**A**) indicates the absent Q coverage due to the gaps between the detectors. It is used as the boundary to separate the two regions defined as low Q region (Q from 0.005 Å$^{-1}$ to 0.026 Å$^{-1}$) and high Q region (Q from 0.031 Å$^{-1}$ to 0.096 Å$^{-1}$). Note that the range of Q covered by the SAXS detectors extends up to 0.095 Å$^{-1}$, and (**A**) shows an enlarged region. (**C**) Difference signal averaged of over the low-Q and the high-Q ranges plotted as functions of the delay time δt. The dashed and dotted lines serve as guides to the eye, highlighting the different trends. (**D**) Change of the integral intensity of the Au (111) Bragg peak as a function of the delay time.

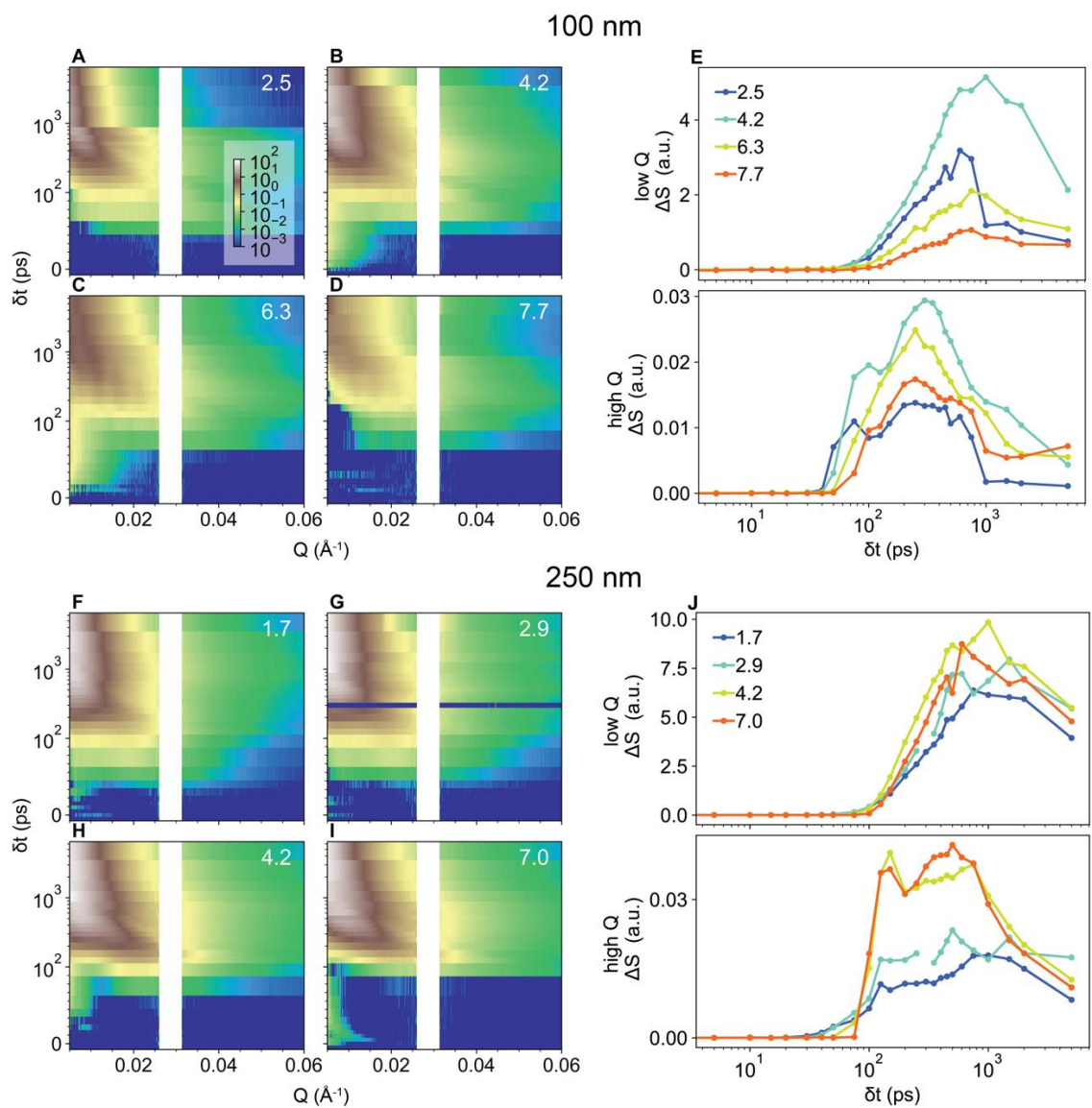

**Fig. 4. Examples of time resolved difference SAXS signal ΔS(Q, δt). (A-D)** Difference SAXS signals measured for 100 nm films irradiated at incident laser fluences of 2.5 J/cm$^2$, 4.2 J/cm$^2$, 6.3 J/cm$^2$, and 7.7 J/cm$^2$. **(F-I)** Difference SAXS signals measured for 250 nm films irradiated at incident laser fluences of 1.7 J/cm$^2$, 2.9 J/cm$^2$, 4.2 J/cm$^2$, and 7.0 J/cm$^2$. The color bar shown in (**A**) is shared by all 8 false color plots. Using the same definition for the low and high Q regions as in Fig. 3, the Q-averaged intensity changes are plotted as functions of the delay time δt in (**E**) for the 100 nm films and in (**J**) for the 250 nm films. Note that the blue row in (**G**) at 300 ps corresponds to an X-ray beam drop during the measurement.

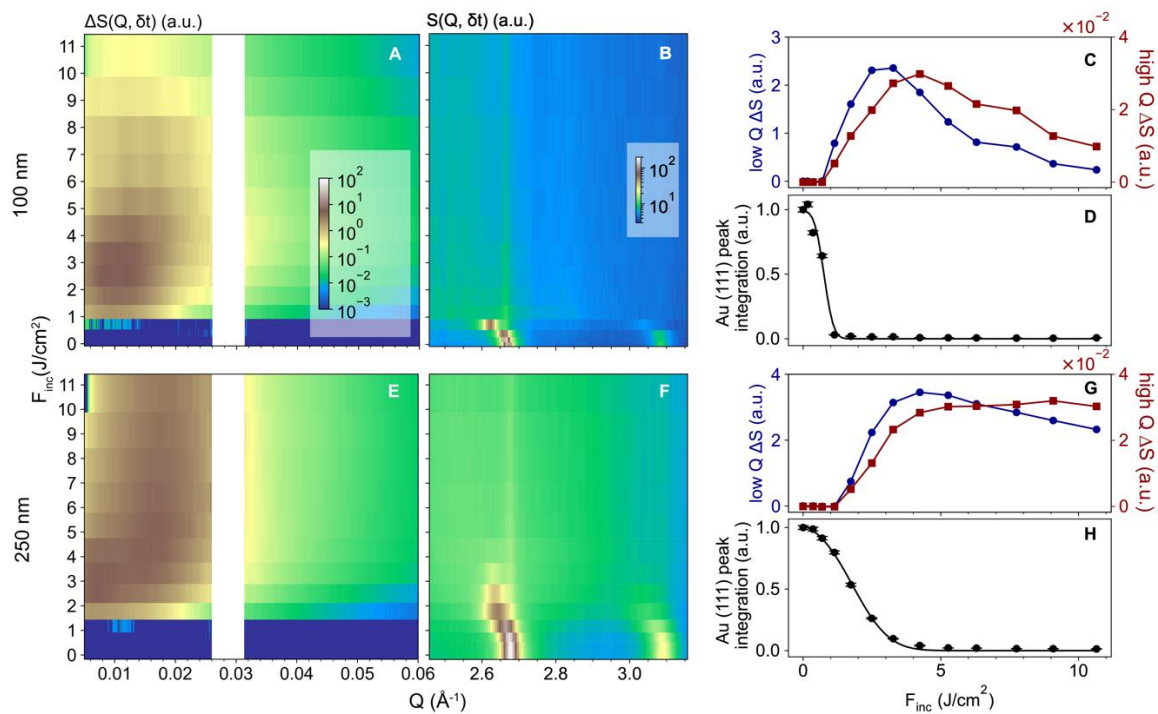

**Fig. 5. Laser fluence dependence.** Radial average of the differential SAXS (**A**) and WAXS (**B**) measured with different laser fluences at a delay $\delta t = 250$ ps for the 100 nm Au film. The corresponding intensity changes in the low and high Q regions, and the Au (1,1,1) Bragg peak integrated intensity are displayed in (**C**) and (**D**). Radial average of the differential SAXS (**E**) and WAXS (**E**) measured with different laser fluences at a delay $\delta t = 200$ ps for the 250 nm Au film. The corresponding intensity changes in the low and high Q regions, and the normalized Au (1,1,1) Bragg peak integrated intensity are displayed in (**G**) and (**H**).

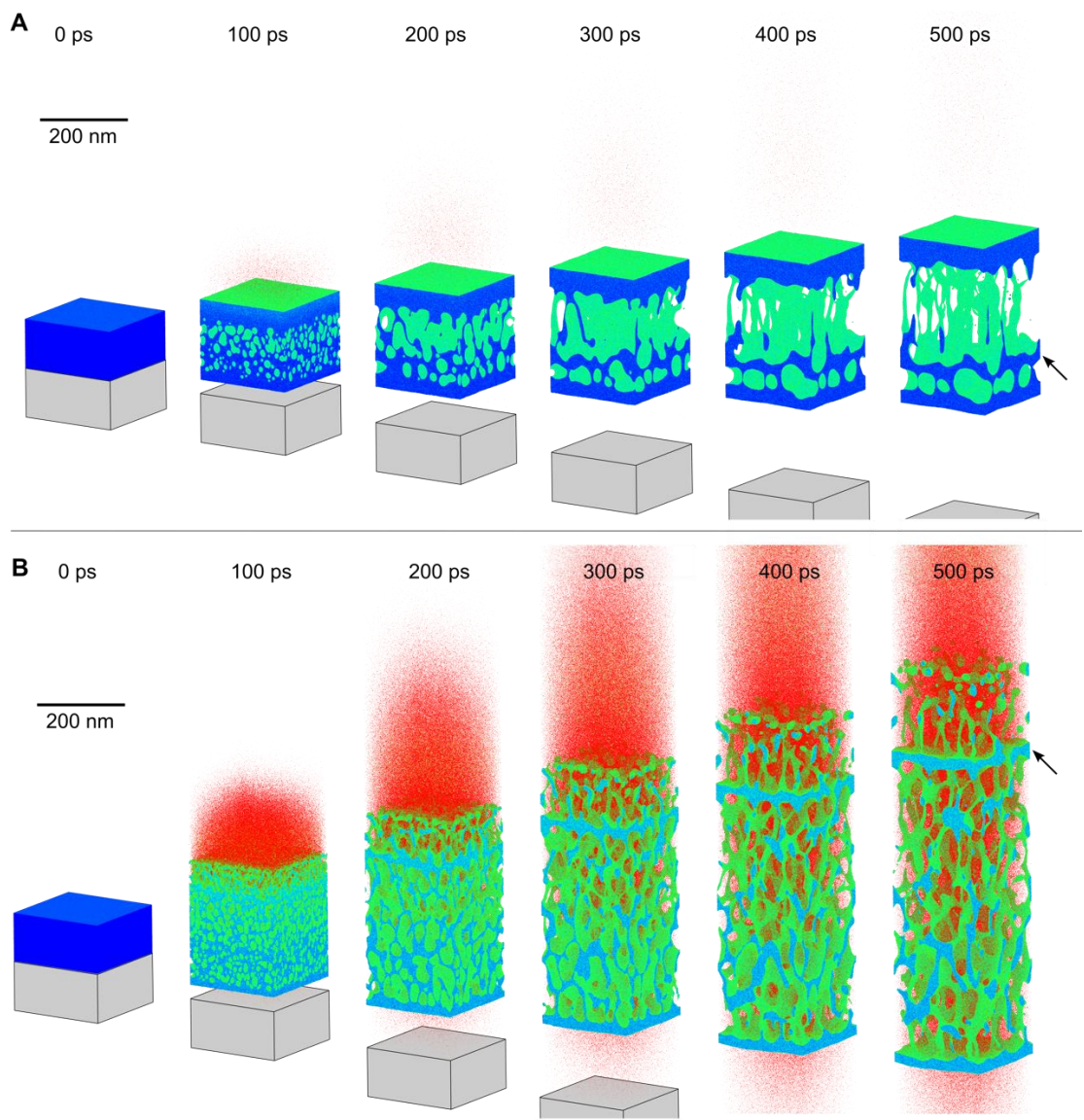

**Fig. 6. Snapshots of atomic configurations from simulations of laser ablation of Au films.** The simulations are performed for 100 nm films deposited on a 100 nm thick $Si_3N_4$ membrane and irradiated by a 50 fs laser pulses. The absorbed laser fluence is 0.1069 J/cm$^2$ in (**A**) and 0.2014 J/cm$^2$ in (**B**). The Au atoms are colored by potential energy in the range of -3.5 to -1 eV, so that the solid and molten Au is blue, surface atoms are green, and the vapor-phase atoms are red. The grey boxes show the position of the substrate at the early stage of the ablation process.

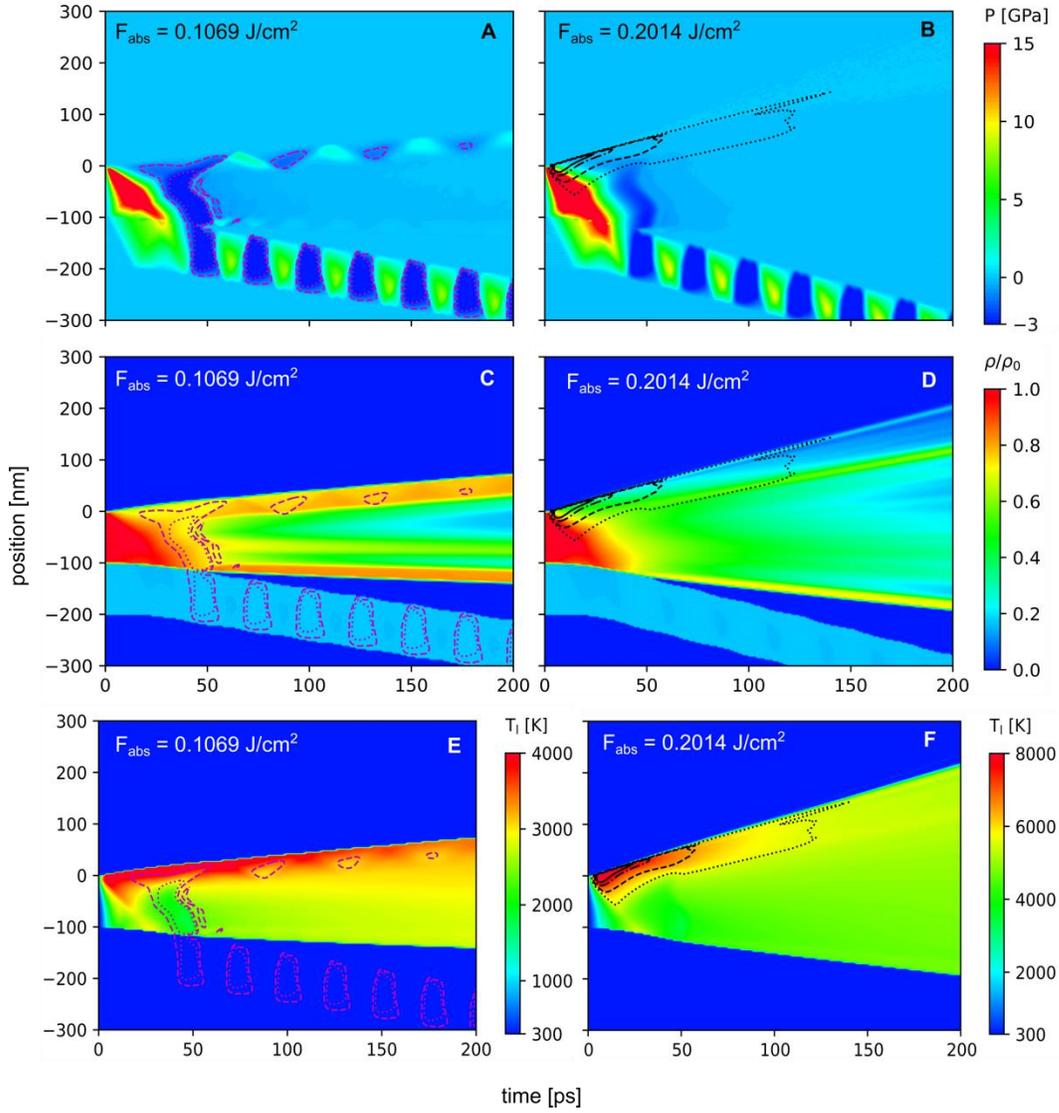

**Fig. 7. Contour plots of pressure, density, and temperature predicted in simulations of laser ablation of Au films.** The simulations are performed for 100 nm films deposited on a 100 nm thick $Si_3N_4$ membrane and irradiated by a 50 fs laser pulses. The absorbed laser fluence is 0.1069 J/cm$^2$ in (**A**, **C**, **E**) and 0.2014 J/cm$^2$ in (**B**, **D**, **F**), and the snapshots of the simulations are shown in Fig. 6. In the contour plots for $F_{abs}$ = 0.1069 J/cm$^2$ (panels **A**, **C**, **E**), the dashed and dotted magenta lines represent the -1 GPa and -3 GPa isobaric lines, respectively. In the contour plots for $F_{abs}$ = 0.2014 J/cm$^2$ (panels **B**, **D**, **F**), the dotted, dashed, dash-dotted and solid lines represent the $0.6T_c$, $0.7T_c$, $0.8T_c$ and $0.85T_c$ isothermal lines, respectively, where $T_c$ is the critical temperature of model Au. The density contour plots (**C**, **D**) are normalized by the room temperature density of Au ($\rho_0$). Since the heat transfer to the $Si_3N_4$ layer is not represented in the simulations, the layer is not colored in the temperature plot. The initial position of the Au film is from 0 to -100 nm, and the substrate is located under the film, from -100 to -200 nm.

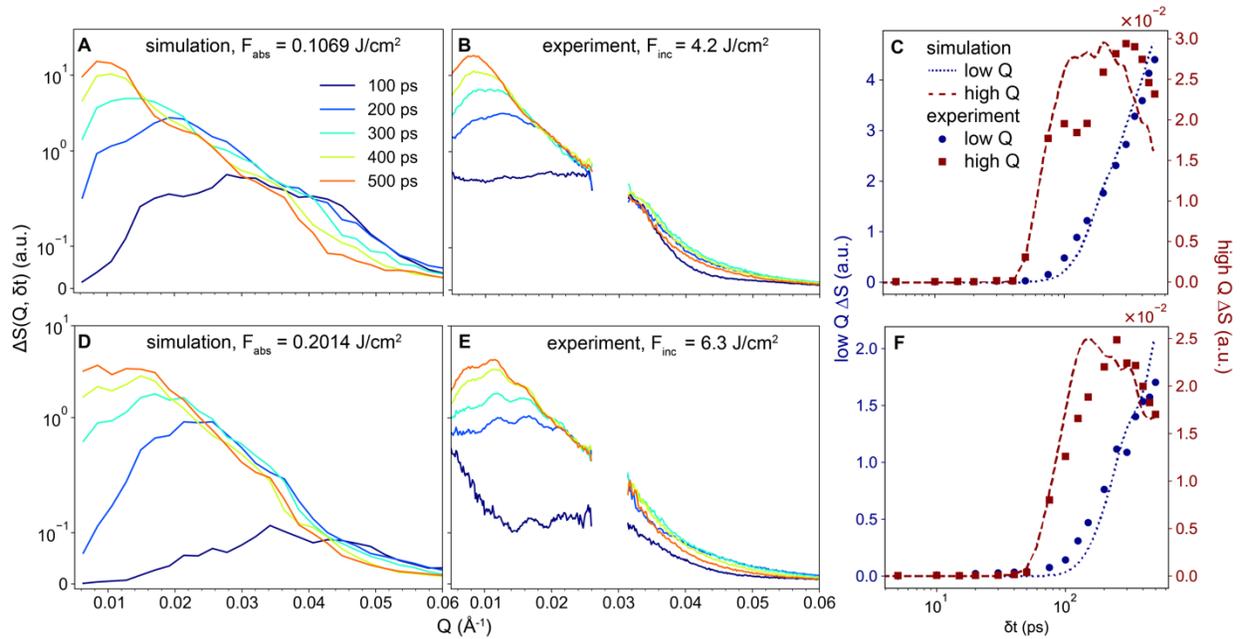

**Fig. 8. Comparison of the simulated and experimental SAXS signals for 100 nm Au films.** The panels in the top row show the radially averaged difference SAXS profile predicted in a simulation at an absorbed laser fluence of 0.1069 J/cm$^2$ (**A**) and measured in experiments at an incident laser fluences of 4.2 J/cm$^2$ (**B**). The corresponding Q-averaged intensity changes in the low and high Q regions are shown in (**C**). The panels in the bottom row show the radially averaged difference SAXS profile predicted in a simulation at an absorbed laser fluence of 0.2014 J/cm$^2$ (**D**) and measured in experiments at an incident laser fluences of 6.3 J/cm$^2$ (**E**). The corresponding Q-averaged intensity changes in the low and high Q regions are shown in (**F**).

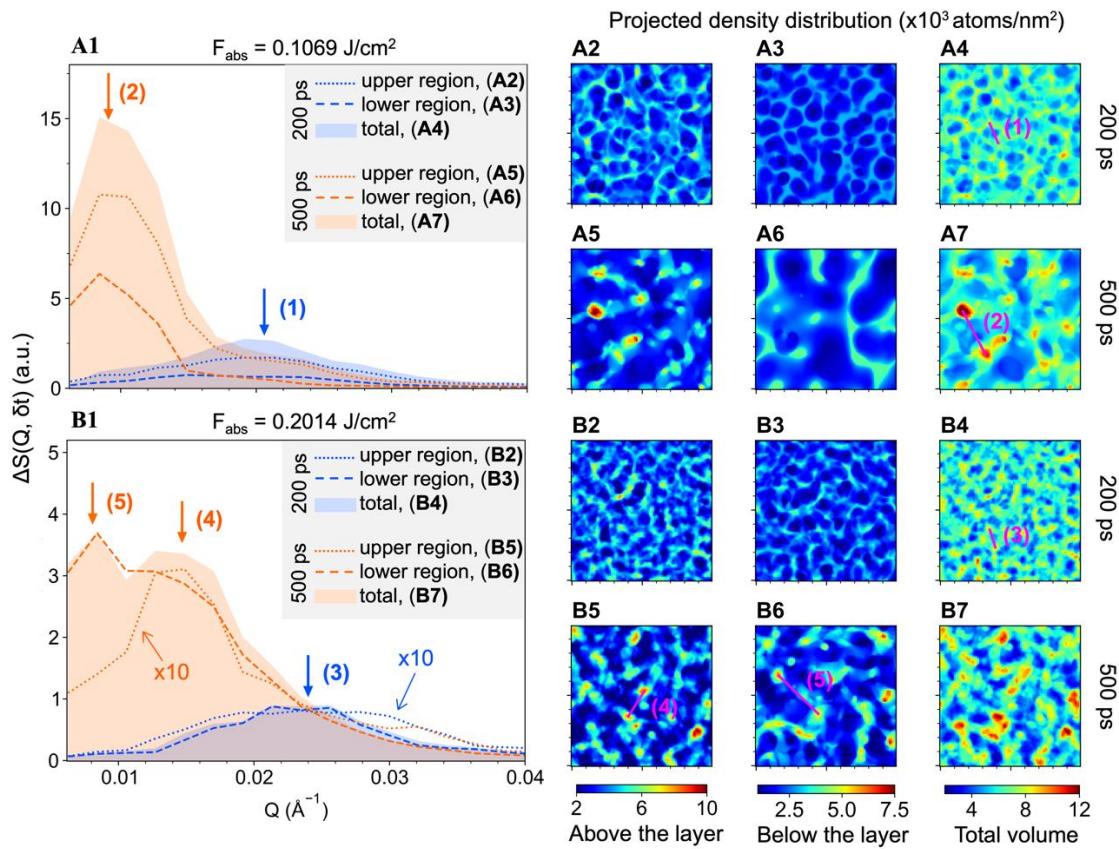

**Fig. 9. Heterogenous structure growth as revealed by the simulation.** The left panels (**A1**) and (**B1**) show the radially averaged difference SAXS intensity profiles predicted for 200 ps and 500 ps after the laser pulse in simulations performed for absorbed laser fluences of 0.1069 J/cm$^2$ and 0.2014 J/cm$^2$. The shaded areas show the total intensity, while the dotted and dashed lines show the contributions to the scattering intensity from material located above and below the corresponding liquid layers marked by the black arrows in Fig. 6. The contribution from above the layer in (**B1**) is magnified by a factor of 10. The projected density distributions in regions above the liquid layer, below the liquid layer, and for the whole system are shown for 200 ps and 500 ps in (**A2**-**A7**) for the lower fluence simulation and in (**B2**-**B7**) for the higher fluence simulation. The blue and red arrows in (**A1**) and (**B1**) correspond to the lengths indicated by the magenta bars shown in (**A4**), (**A7**), (**B4**), (**B5**), and (**B6**). The size of the simulation box in (**A2**-**A7**) and (**B2**-**B7**) is 200 × 200 nm$^2$.

# Supplementary Materials for

## Dynamics of Nanoscale Phase Decomposition in Laser Ablation

**Authors**

Yanwen Sun[1*‡], Chaobo Chen[2‡], Thies J. Albert[3], Haoyuan Li[1], Mikhail I. Arefev[2], Ying Chen[1], Mike Dunne[1], James M. Glownia[1], Matthias Hoffmann[1], Matthew J. Hurley[4,7], Mianzhen Mo[1], Quynh L. Nguyen[1], Takahiro Sato[1], Sanghoon Song[1], Peihao Sun[5], Mark Sutton[6], Samuel Teitelbaum[7], Antonios S. Valavanis[2], Nan Wang[1], Diling Zhu[1], Leonid V. Zhigilei[2*], Klaus Sokolowski-Tinten[3*]

**Affiliations**

[1]Linac Coherent Light Source, SLAC National Accelerator Laboratory, USA.

[2]Department of Materials Science and Engineering, University of Virginia

[3]Department of Physics, Universität Duisburg-Essen, Germany.

[4]Department of Physics, Stanford University, USA.

[5]Department of Physics, Università degli Studi di Padova, Italy.

[6]Department of Physics, McGill University, Canada.

[7]Department of Physics, Arizona State University, USA.

[‡] These authors contributed equally to this work

[*]To whom correspondence should be addressed; E-mail: yanwen@slac.stanford.edu, lz2n@virginia.edu, klaus.sokolowski-tinten@uni-due.de.

CONTENTS

## 1. Characterization of polycrystalline Au films

Using $l$ to denote the grain size of the polycrystalline gold film and $D(l)$ as the size distribution function, the scattering intensity can be written as

$$I(Q) \propto \int_0^\infty D(l)|\mathcal{F}(Q,l)|^2 dl. \tag{S1}$$

Here $|\mathcal{F}(Q,l)| = 3J_1(Ql/2)/(Ql/2)$ with $J_1$ being the Bessel function of the first kind of order 1. Assuming a Gaussian distribution for the grains. With an average size of $l_0$ and a standard deviation $\sigma_l$:

$$D(l) = \frac{1}{\sigma_l\sqrt{2\pi}} \exp\left[\frac{-(l-l_0)^2}{2\sigma_l^2}\right]. \tag{S2}$$

By fitting the radial distribution of the SAXS profile from the unexcited Au film using the above equation, we get an averaged grain size of $52 \pm 8$ nm. This is in good agreement with atomic force microscope (AFM) and scanning electron microscope (SEM) measurements as shown in Fig. S1 (**B**, **C**).

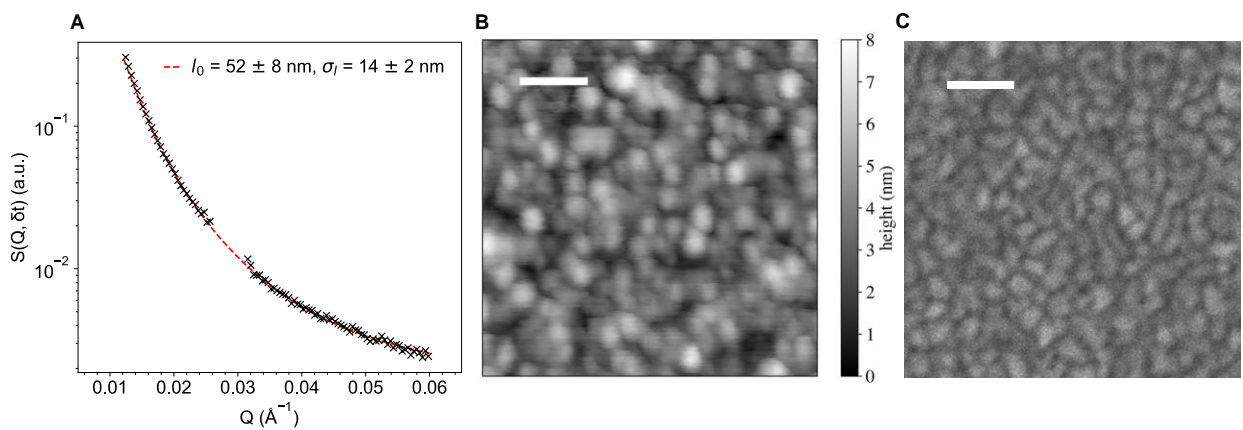

**Fig. S1. Characterization of the polycrystalline Au films.** (**A**) SAXS examination of the unexcited 100 nm Au film. The cross marker denotes the radial average of the SAXS background displayed in Fig. 2 (second row, first image) in the main paper and the red curve represents the fit to Equation **S1** yielding the averaged grain size. (**B**) AFM and (**C**) SEM measurements on the 100 nm Au film. The scale bar in both plots denote 100 nm.

## 2. Analysis of the WAXS data

Fig. S2 displays the time resolved WAXS measurement of the 100 nm Au film before and after laser excitation with an incident fluence of 10.5 J/cm$^2$. The parasitic scattering background was approximated using a linear fit. The Au (111) peak was modelled using a Gaussian distribution, allowing for the calculation of the area under the peak. The fitting routine was utilized to plot Fig. 3 (D) and Fig. 5 (D, H).

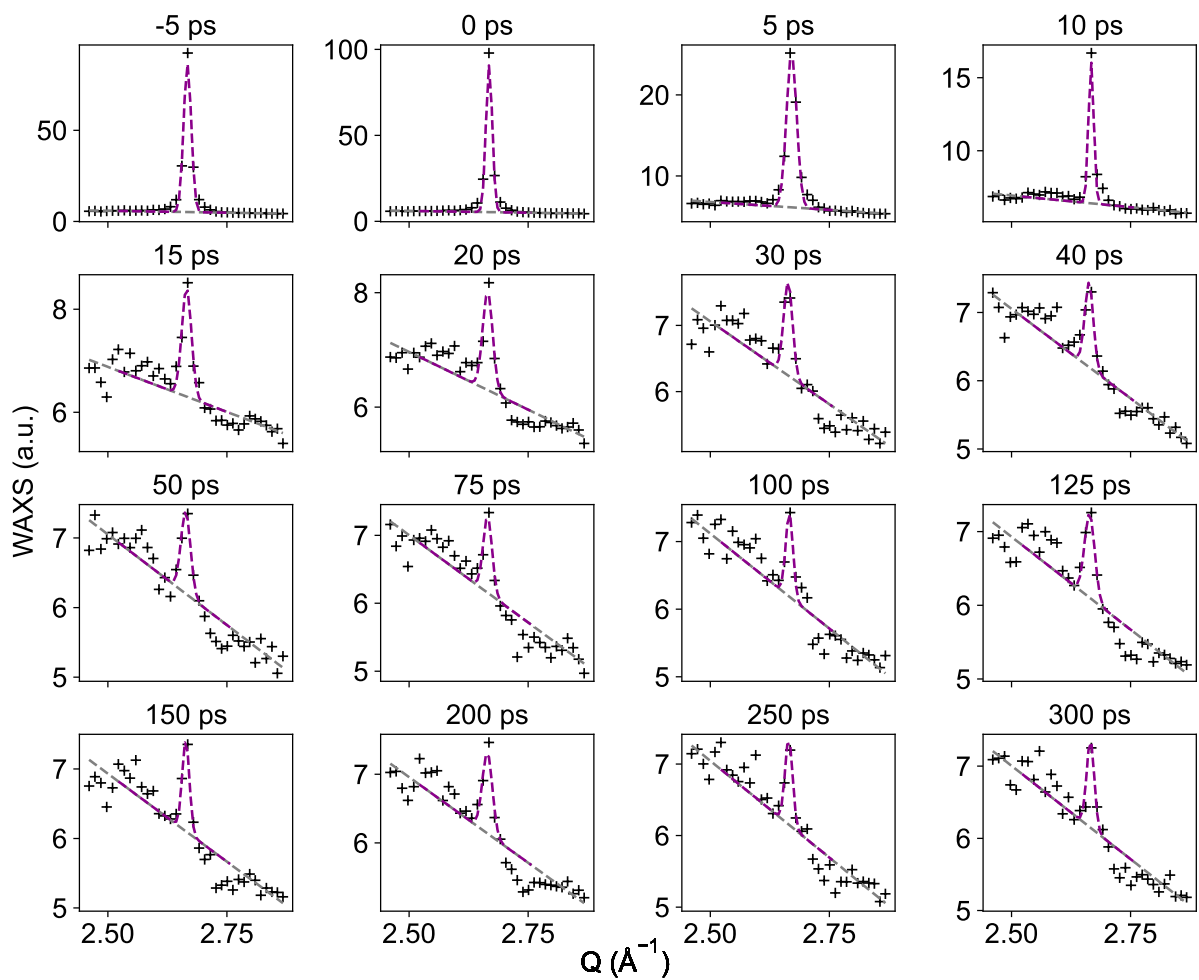

**Fig. S2. Fitting of the Au (111) peak to obtain integral peak intensity.** The WAXS data denoted by crosses was fitted using two components: (1) a linear term accounting for the parasitic scattering background (dashed gray line) and (2) a Gaussian fit to the Au (111) peak. The magenta line shows the sum of the two components.

## 3. Computational model for laser interaction with Au films: Solution of electromagnetic wave equations (EM) coupled with two-temperature model (TTM).

The interaction of a femtosecond laser pulse with an Au film is represented in the present study by a model combining the solution of electromagnetic wave equations (EM) with the two-temperature model (TTM). Within the coupled TTM-EM model, the TTM accounts for the electron heat transport in the Au film during the laser energy deposition, providing the electron and lattice temperature fields for the electromagnetic wave calculation. The TTM model incorporates temperature-dependent electron heat capacity and electron-phonon coupling constant, determined through *ab-initio* calculations (*1*). The electron thermal conductivity $k_e$ is modeled using the Drude model relationship, $k_e(T_e, T_l) = 1/3\, v_f^2 C_e \tau_e$, where $v_f$, $C_e$, and $\tau_e$ are the Fermi velocity, electron heat capacity, and electron relaxation time. The parameters for the heat conductivity expression are the same as those used in Ref. (*2*).

The EM modeling involves a numerical solution of Maxwell equations using a finite-difference time-domain scheme proposed by Yee (*3*). The optical properties within the computation domain are defined by the constitutive relation between electric field intensity ***E*** and electric displacement, $\boldsymbol{D} = \varepsilon_0\, \varepsilon_r(\omega) \boldsymbol{E}$. The relative permittivity of vacuum ($\varepsilon_r = 1$) and $Si_3N_4$ ($\varepsilon_r = 4.0974$ for 800 nm laser (*4*)) are assumed to be constant. The permittivity of the Au film is modeled using the Drude model with the plasmonic frequency and the electron relaxation time fitted to the room temperature experimental values (*5*) of refractive index and extinction coefficient at 800 nm. The temperature dependence of the electron relaxation time is described by a common expression, $1/\tau = AT_e^2 + BT_l$ (*6*), with $A = 3.043 \times 10^6$ Hz/K$^2$, $B = 4.7159 \times 10^{11}$ Hz/K, where $A$ value is the same value for thermal conductivity (*2*) and $B$ is fitted to reproduce the refractive index (*5*) at room temperature. The Drude model assumes that the response of Au to the electromagnetic wave is solely defined by the density of free electrons and the electron relaxation time, while the contributions from both free and bonded electrons should be considered for a more accurate representation of optical properties of Au under conditions of strong optical excitation (*7,8*). While the approximate nature of this model prevents the quantitative conversion of F$_{inc}$ to F$_{abs}$ at high fluences, the TTM-EM model still provides a more accurate description of the laser energy deposition as compared to the use of the Beer-Lambert law with constant room temperature values of the reflectivity and absorption coefficient. The Drude model also predicts the reflectivity drop with increasing electron temperature (*9*), although this drop is likely to be overestimated (*10*, *11*).

The laser beam in the TTM-EM model is represented by a planar laser source positioned above the Au film and emitting electromagnetic waves. The perfectly matched layer (PML) boundary condition (*12*) is applied at the top and bottom of the TTM-EM simulation domain. The PML boundary fully absorbs the electromagnetic wave reflected from the Au film, which mimics the wave propagation into an infinitely large vacuum space above the irradiated film. The deposited energy density, *e*, (in units of J/cm$^3$) is calculated by Joule's law, $e = \int_0^\tau \boldsymbol{j} \cdot \boldsymbol{E} dt$, where $\tau, \boldsymbol{j}$, and $\boldsymbol{E}$ are the duration of the simulation, current density, and electric field, respectively.

## 4. Electromagnetic modeling of the effect of surface roughness on laser interaction with polycrystalline Au films

In order to evaluate the effect of the surface roughness of polycrystalline Au films on the laser energy deposition, we create a model system mimicking the surface topography measured by AFM (see supplementary Section 1). We then apply the TTM-EM model described in the supplementary Section 3 to predict the lateral distribution of the laser energy deposition in this model system. The TTM-EM simulation is performed for a 200 nm × 200 nm part of the film outlined by the dashed square in the experimental AFM image in Fig. S3 (**A**). The computational system is exposed to a 50 fs laser pulse with a linear polarization, as indicated by the arrow in the figure. The local absorbed fluence is calculated by integration of the deposited energy density *e* over the depth of the film and is shown in Fig. S3 (**B**). The average absorbed fluence of the laser pulse is 0.2863 J/cm$^2$, with a standard deviation of 0.0220 J/cm$^2$ describing the distribution of fluence along the film surface.

The distribution of the absorbed fluence exhibits a nanoscale pattern of reduced and enhanced absorption spots with characteristic sizes in tens of nanometers. The spots tend to be elongated along the *x*-axis, *i.e.*, perpendicular to the polarization direction. The temperature-dependent optical properties amplify the heterogeneity of laser energy deposition, providing a positive feedback to the absorption enhancement (*13*). While the electron heat transfer acts to redistribute the deposited energy along the film surface, the electron and lattice temperature distributions shown in Fig. S3 (**C**) and (**D**) indicate that the uneven energy deposition can produce a substantial temperature variation along the film surface. The distributions are plotted for 2 ps, after the end of the laser pulse but prior to the completion of electron-phonon equilibration. As a result, the electron temperature in Fig. S3 (**C**) is still substantially higher than the lattice one plotted in Fig. S3 (**D**).

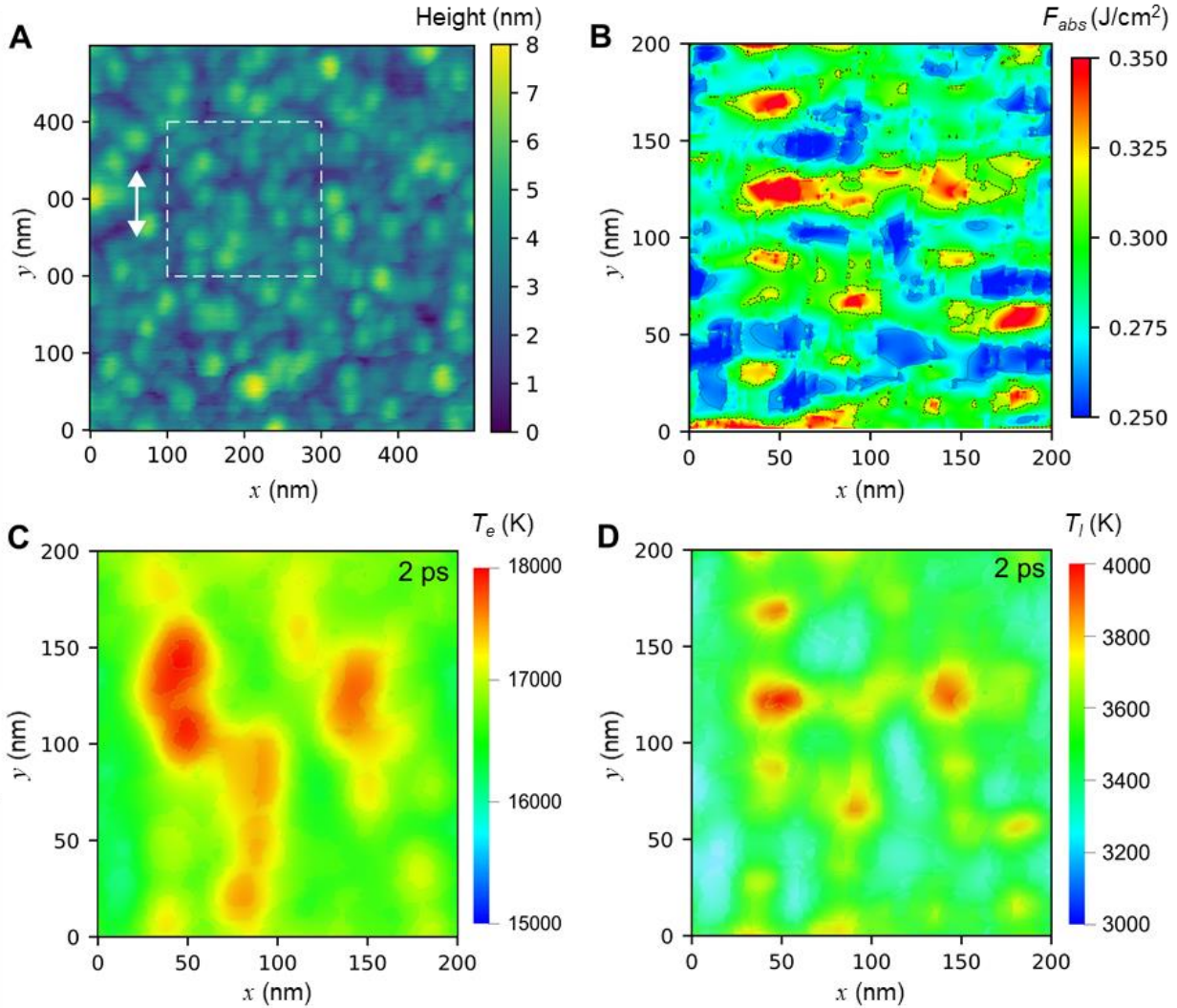

**Fig. S3. Lateral heterogeneity of the laser energy deposition on a rough film surface.** (**A**): surface roughness of a 100 nm Au film imaged by AFM, also shown in Fig. S1 (**A**). The white dashed square outlines the region used for the preparation of a digital sample for the TTM-EM simulation. The white arrow shows the direction of laser light polarization assumed in the simulation. (**B**): The variation of the absorbed laser fluence along the film surface predicted in the TTM-EM simulation. The dashed and dotted lines are isolines for the values of the mean fluence

plus/minus the standard deviation. (**C**) and (**D**): The distributions of the electron (**C**) and lattice (**D**) temperatures along the film surface predicted in the TTM-EM simulation for a time of 2 ps.

The formation of the hot spots may result in the variation of the ablation characteristics along the surface of the film, leading to the lateral modulation of density in the ablation plume, as has been suggested by earlier atomistic simulations of the ablative LIPSS formation (*14*, *15*). This modulation could be the origin of the rapid emergence of the SAXS signals in the low-Q region observed in the experimental profiles for the probe delays of 100 ps, Fig. 8 (**B**, **E**). Indeed, the "hot" and "cold" spots in Fig. S3 (**D**) are separated from each other by relatively large distances, suggesting that the spatially modulated ablation may result in a prompt appearance of a SAXS signal in the low Q region. The phenomenon of uneven energy deposition is expected to be more pronounced in targets with greater roughness, lower thermal conductivity, and stronger electron-phonon coupling. Note that the experimental SAXS patterns in Figs. 1 and 2 do not exhibit any noticeable anisotropy in the polarization direction, suggesting that the high thermal conductivity of Au results in the smearing of any anisotropy in the laser energy deposition during the time of the electron-phonon equilibration.

## 5. Energy density required for complete melting and phase explosion (*evaluation with experimental parameters for Au*)

It is instructive to relate the energy deposited by the laser pulse to the energy density required for causing phase transformations that can be detected in WAXS and SAXS probing experiments. The conversion of the absorbed laser fluence to the energy density can be done by assuming uniform energy distribution throughout the thickness of the irradiated film. As can be seen from the temperature contour plots in Fig. 7 (**E**, **F**), the laser energy deposition generates a substantial temperature gradient within the film. Still, the evaluation of the average energy density provides a useful reference for analysis of the experimental and computational results. In particular, the energy density required for the complete melting of the uniformly heated film can be considered to be the lower bound estimate for the complete melting under realistic energy distribution in the film. The absorbed fluence that corresponds to the energy density required for uniformly heating the film up to the limit of thermodynamic stability against the explosive decomposition into vapor and liquid droplets, $T^* = 0.9T_c$ (*16-18*), can be considered the upper bound estimate of the threshold for the transition to the phase explosion regime of the film disintegration.

The reference levels of energy are evaluated as follows:

$E^S_{T_m} = \int_{300}^{T_m} C^s_p(T)\, dT$ - the energy needed for heating solid Au up to its melting temperature $T_m$,

$E^l_{T_m} = E^S_{T_m} + \Delta H_m$ - the energy needed for complete melting, where $\Delta H_m$ is the heat of melting,

$E^l_{T^*} = E^l_{T_m} + \int_{T_m}^{T^*} C^l_p(T)\, dT$ - the energy needed for superheating the molten Au up to the limit of thermodynamic stability against the explosive phase decomposition $T^*$. Note that this quantity is evaluated by extending the integration beyond the equilibrium boiling temperature at 1 atm pressure. As discussed theoretically by Miotello and Kelly (*16*), the buildup of the saturated vapor pressure above the irradiated surface and the nucleation and growth of vapor bubbles in normal boiling are kinetically limited under conditions of rapid laser heating, and the molten Au can be superheated up about $0.9T_c$, when the release of vapor proceeds in an explosive manner, in a process commonly referred to as "phase explosion" (*16-18*).

The reference levels of energy are calculated using the experimental parameters of Au. The temperature dependence of the lattice heat capacity predicted for a temperature range from 300 K to $T_m = 1337$ K can be approximated as $C_p^s(T) = \alpha\,T^{-2} + \beta + \gamma T + \delta T^2$, where $\alpha$ = -37708.8 J K mol$^{-1}$, $\beta$ = 25.766 J mol$^{-1}$ K$^{-1}$, $\gamma$ = -1.715184×10$^{-3}$ J mol$^{-1}$ K$^{-2}$, and $\delta$ = 4.9527×10$^{-6}$ J mol$^{-1}$ K$^{-3}$ (*19*). For the molten Au, from $T_m$ to $T^*$, the heat capacity is approximated by a constant value, $C_p^l$ = 32.97 J mol$^{-1}$ K$^{-1}$ (*19*). The enthalpy of melting is $\Delta H_m$ = 12.72 kJ mol$^{-1}$ (*19*). The temperature for the onset of the phase explosion, $T^* = 0.9T_c = 6660$ K, is obtained with $T_c = 7400$ K (*20*).

With these parameters, the energy density required for heating the film to $T_m$ and fully melting it at $T_m$ is $E_{T_m}^l$ = 41.8 kJ mol$^{-1}$. The energy density for heating the film from 300 K to $T^*$ is $E_{T^*}^l = E_{T_m}^l + 175.5$ kJ mol$^{-1}$ = 217.3 kJ mol$^{-1}$. Assuming a uniform distribution of the deposited energy throughout the thickness of a 100 nm Au film, the values of the deposited energy density can be converted to the corresponding values of the absorbed laser fluence, 0.0410 J cm$^{-2}$ for $E_{T_m}^l$ and 0.213 J cm$^{-2}$ for $E_{T^*}^l$.

## 6. Energy density required for complete melting and phase explosion (*evaluation with parameters for model EAM Au*)

Similar to the calculations of the values of energy density performed in the previous section with experimental parameters for Au, it is useful for interpretation of the computational results to calculate the corresponding values with the thermodynamic properties predicted by the Embedded Atom Method (EAM) interatomic potential (*21*) used in the simulations.

The thermodynamic parameters predicted with the EAM potential are the melting temperature $T_m^{EAM} = 1318\ K$, the thermodynamic critical temperature $T_c^{EAM} = 9250\ K$ (21) and the latent heat of melting $\Delta H_m^{EAM}$ = 13.03 kJ mol$^{-1}$. The temperature dependence of the lattice heat capacity, predicted in a simulation of heating of an Au crystal from 300 K to $T_m^{EAM}$ at a constant 1 atm pressure, can be approximated by the following expression, $C_p^{s,EAM}(T) = \alpha + \beta T + \gamma T^2$, where $\alpha$ = 25.26 J mol$^{-1}$ K$^{-1}$, $\beta$ = -1.253×10$^{-3}$ J mol$^{-1}$ K$^{-2}$, and $\gamma$ = 4.399×10$^{-6}$ J mol$^{-1}$ K$^{-3}$. Beyond the melting point, from $T_m^{EAM}$ to $0.9T_c^{EAM}$, the temperature dependence of the heat capacity is found to be well described by a linear dependence $C_p^{l,EAM}(T) = \alpha' + \beta'T$, where $\alpha'$ = 24.43 J mol$^{-1}$ K$^{-1}$, $\beta'$ = 87.79×10$^{-3}$ J mol$^{-1}$ K$^{-2}$. The temperature-dependent electronic contribution to the total heat capacity of the material (1) is also accounted for in the TTM-MD model, yielding the total heat capacity (vibrational plus electronic) ranging from 25.5 J mol$^{-1}$ K$^{-1}$ at 300 K to 44.97 J mol$^{-1}$ K$^{-1}$ at temperature approaching $0.9T_c^{EAM}$. Note that an onset of explosive phase decomposition of the computational system (rapid spontaneous increase in the volume of the system at 1 atm pressure) was observed at 8300 K, *i.e.*, at $0.897T_c^{EAM}$. Given the stochastic nature of the phase decomposition onset and its dependence on the size of the system and the heating rate (the heating rate of 50 K/ps used in the simulation is comparable to but slower than the heating expected due to the electron-phonon coupling in Au excited by a laser pulse in the ablation regime), we extrapolated the temperature dependence of the heat capacity and extended the integration limit to $0.9T_c^{EAM}$.

With these parameters, the energy density required for the complete melting of the model EAM Au material at $T_m^{EAM}$ = 1318 K is calculated as $E_{T_m}^{l,EAM}$ = 41.6 kJ mol$^{-1}$. The energy density for heating Au up to $T^{*,EAM} = 0.9T_c^{EAM}$ = 8325 K is calculated as $E_{T^*}^{l,EAM}$ = 285.7 kJ mol$^{-1}$. Assuming a uniform distribution of the deposited energy throughout the thickness of a 100 nm Au film, the values of the deposited energy density can be converted to the corresponding values of the absorbed laser fluence, 0.0408 J cm$^{-2}$ for $E_{T_m}^{l,EAM}$ and 0.280 J cm$^{-2}$ for $E_{T^*}^{l,EAM}$.

## 7. Beam size characterization for the optical laser

The size of the laser beam was characterized by employing the Liu method (22), which assesses the fluence dependence of the area modified by the laser, as detailed below.

The incident laser has a (close to) Gaussian profile following $F(r) = F_0 \cdot \exp[-(r/r_0)^2]$, and the pulse energy can therefore be written as $E_0 = \pi r_0^2 F_0$. Given that the surface modification has a fluence threshold corresponding to $F_{th}$, this then gives the area of the surface modification $A_{mod}$ with:

$$\ln(E_0) - \ln(E_{th}) = \frac{1}{\pi r_0^2} \cdot A_{mod}. \tag{S3}$$

Here, the threshold pulse energy $E_{th}$, corresponding to $A_{mod} = 0$, can be expressed as $E_{th} = \pi r_0^2 F_{th}$.

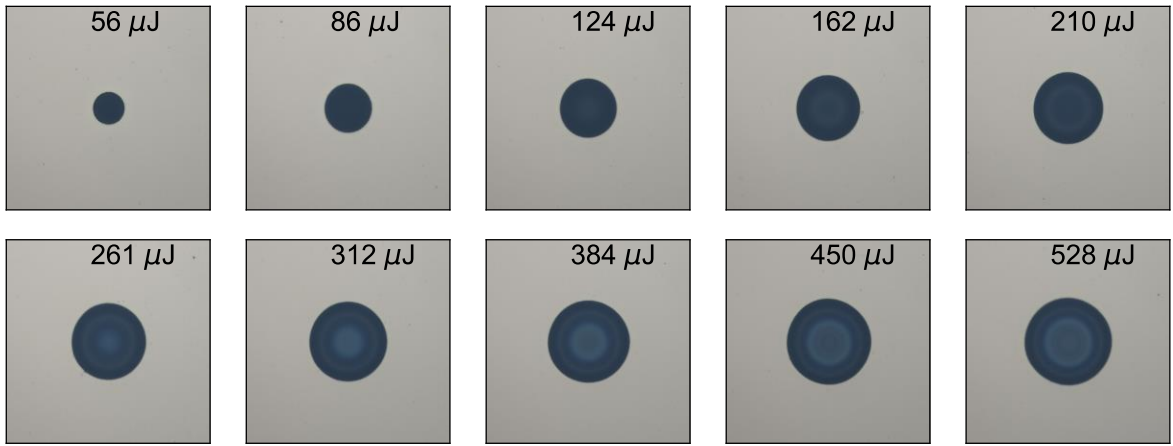

**Fig. S4. Microscope image of the laser imprints at different incident laser fluences.** The field of view of each image is 300×300 $\mu m^2$. The incident laser pulse energy is indicated in the upper right corner of each image.

The measurement shown in Fig. S4 was performed on a 50 nm Au film deposited on the silicon frame (thickness 200 $\mu m$) of the used $Si_3N_4$ membrane arrays, where the incident laser pulse energies were varied. The modified areas were extracted and plotted against the corresponding $\ln(E_0)$, as displayed in Fig. S5. A linear fit to Equation (S3) determines the beam full width at half maximum (FWHM) to be $d_{FWHM} = (66 \pm 0.4\ \mu m)$, $E_{th} = (39 \pm 3\ \mu J)$, and the corresponding threshold fluence $F_{th} = (0.79 \pm 0.07)$ J/cm$^2$.

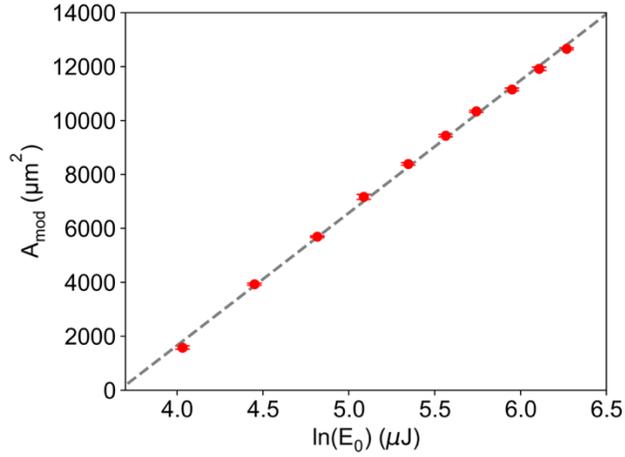

**Fig. S5. Beam size characterization of the optical laser using the Liu method.** The error of the modified area $A_{mod}$ is estimated using the standard deviation of 6 imprints. Dashed gray line corresponds to a linear fit to Equation (S3) and gives a laser FWHM of $d_{FWHM} = (66 \pm 0.4\ \mu m)$, $E_{th} = (39 \pm 3\ \mu J)$, and $F_{th} = (0.79 \pm 0.07)$ J/cm$^2$.

## 7. References for the Supplementary Materials